\documentclass[fleqn,usenatbib]{mnras}

\usepackage{newtxtext,newtxmath}
\usepackage{todonotes}
\usepackage{subcaption}
\usepackage{hyperref}

\newcommand{\Msun}{\ensuremath{\mathrm{M}_\odot}}
\newcommand{\Fable}{\textsc{Fable}}

\usepackage[normalem]{ulem}

\usepackage[T1]{fontenc}

\DeclareRobustCommand{\VAN}[3]{#2}
\let\VANthebibliography\thebibliography
\def\thebibliography{\DeclareRobustCommand{\VAN}[3]{##3}\VANthebibliography}

\usepackage{graphicx}	
\usepackage{amsmath}	

\title[Premature SMBH mergers in \Fable{}]{Premature supermassive black hole mergers in cosmological simulations of structure formation}

\author[S. Buttigieg et al.]{
Stephanie Buttigieg,$^{1,2}$\thanks{E-mail: sb2583@cam.ac.uk (SB)}
Debora Sijacki,$^{1,2}$
Christopher J.\ Moore $^{1,2,3}$ and 
Martin A. Bourne$^{1,2,4}$
\\
$^{1}$Institute of Astronomy, University of Cambridge, Madingley Road, Cambridge CB3 0HA, UK\\
$^{2}$Kavli Institute for Cosmology (KICC), University of Cambridge, Madingley Road, Cambridge CB3 0HA, UK\\
$^{3}$Department of Applied Mathematics and Theoretical Physics, Centre for Mathematical Sciences, University of Cambridge, \\ Wilberforce Road, Cambridge, CB3 0WA, UK\\
$^{4}$Centre for Astrophysics Research, Department of Physics, Astronomy and Mathematics, University of Hertfordshire, \\ College Lane, Hatfield, AL10 9AB, UK
}

\date{MNRAS, accepted}

\pubyear{\the\year{}}

\begin{document}
\label{firstpage}
\pagerange{\pageref{firstpage}--\pageref{lastpage}}
\maketitle

\begin{abstract}
The co-evolution of massive black holes (BHs) and their host galaxies is well-established within the hierarchical galaxy formation paradigm. Large-scale cosmological simulations are an ideal tool to study the repeated BH mergers, accretion and feedback that conspire to regulate this process. While such simulations are of fundamental importance for understanding the complex and intertwined relationship between BHs and their hosts, they are plagued with numerical inaccuracies at the scale of individual BH orbits. To quantify this issue, taking advantage of the $(100 \, h^{-1}\,\text{cMpc})^3$ \Fable{} simulation box, we track all individual BH mergers and the corresponding host galaxy mergers as a function of cosmic time. We demonstrate that BH mergers frequently occur prematurely, well before the corresponding merger of the host galaxies is complete, and that BHs are sometimes erroneously displaced from their hosts during close galaxy encounters. Correcting for these artefacts results in substantial \textit{macrophysical delays}, spanning over several Gyrs, which are additional to any \textit{microphysical delays} arising from unresolved BH binary hardening processes. We find that once the \textit{macrophysical delays} are accounted for, high-mass BH merger events are suppressed, affecting the predictions for the BH population that may be observable with \textit{LISA} and pulsar timing arrays. Furthermore, including these macrophysical delays leads to an increase in the number of observable dual active galactic nuclei, especially at lower redshifts, with respect to \Fable{}. Our results highlight the pressing need for more accurate modelling of BH dynamics in cosmological simulations of galaxy formation as we prepare for the multi-messenger era.
\end{abstract}

\begin{keywords}
black hole mergers -- quasars: supermassive BHs -- methods: numerical -- galaxies: kinematics and dynamics -- gravitational waves
\end{keywords}



\section{Introduction} \label{sec:introduction}

Observations provide compelling evidence that most, if not all, massive galaxies host a central supermassive black hole (SMBH) with masses ranging from $10^6$ \Msun{} to $10^{11}$ \Msun{}~\citep{magorrian_1998, richstone_1998, Kormendy_2013}. Furthermore, these observations suggest a relation between SMBHs and their host galaxies, with the SMBH mass being found to correlate with key stellar properties, such as bulge luminosity, mass, and velocity dispersion~\citep{magorrian_1998, gebhardt_2000, Ferrarese_2002, Termaine_2002, Marconi_2003, haring_2004, gultekin_2009, Kormendy_2013, McConnell_2013}. These relations suggest an evolutionary link between SMBHs and their host galaxies, influenced by processes such as galaxy mergers, active galactic nuclei (AGN) and stellar feedback~\citep{dimatteo_2005, Sijacki_2007, hopkins_2008, somerville_2008,dubois_2016, Terrazas_2020}. However, significant uncertainties remain in these scaling relations and their evolution across cosmic time~\citep{Merloni_2009}, with observations from the James Webb Space Telescope (\textit{JWST}) starting to provide better constraints on these relations at higher redshifts~\citep[see e.g.][]{Maiolino_2024}.

In the $\Lambda$CDM model, galaxies are expected to undergo mergers throughout the history of the Universe. When both galaxies involved in a merger host a SMBH, it is expected that a SMBH binary will form, which may eventually lead to coalescence. Before the two merging galaxies coalesce into a single system, it is the galaxy merger process itself that drives the shrinking of the SMBH separation, as the black holes (BHs) remain gravitationally bound to the centres of their respective host galaxies. Once the galaxies merge to form a unified remnant, the continued orbital decay of the SMBH binary is governed by a series of physical processes operating across different time and length scales, as outlined in the seminal work of \citet{begelman_1980}.

During the galaxy merger, the two BHs are still an unbound pair and migrate toward the centre of the host galaxy via dynamical friction (DF)~\citep{chandrasekhar_1943, binney_2011}. As a BH moves through a background of gas, stars, and dark matter (DM), it generates a wake that exerts a gravitational drag force on the BH, causing it to slow down~\citep{chandrasekhar_1943, antonini_2012, Dosopoulou_2017}. When the binding energy of the BH pair exceeds the average kinetic energy of the surrounding stars, the system transitions into a `hard' binary~\citep{binney_2011}. At this stage, DF becomes inefficient, and one of the primary mechanisms for orbital energy extraction is stellar scattering. Stars with sufficiently low angular momentum, residing in the so-called `loss cone' of parameter space, interact with the binary. These interactions transfer energy from the binary to the stars, ejecting them at high velocities and shrinking the binary orbit in the process~\citep{Quinlan_1996, Khan_2013}. Additionally, in gas-rich environments, a circumbinary disc (CBD) can exert gravitational torques on the BHs, facilitating orbital decay at separations of approximately 0.1–10 pc~\citep{pringle_1991, artymowicz_1994, gould_2000, Armitage_2005, lodato_2009}, with gas residing within the ``cavity'' of the CBD, such as streams and mini-discs, being shown to have significant impact in this regime~\citep{Franchini_2022, siwek_2023, bourne_2024}. At sufficiently small separations ($\lesssim 10^{-3}$ pc), gravitational wave (GW) emission becomes the dominant mechanism for energy and angular momentum loss, driving the binary to coalescence~\citep{Peters_1964}. Throughout these hardening mechanisms, the BHs may continue to accrete material, leading to changes in their mass and spin as they evolve toward their final pre-merger states~\citep{bardeen_1975, King_1999, merritt_2005, Gerosa_2015, Siwek_2020, bourne_2024}.

\textit{JWST} is revolutionising our understanding of the high-redshift SMBH population, enabling detailed studies of the redshift evolution of SMBH–host galaxy relationships out to redshifts as high as $z\approx 10$~\citep{Pacucci_2023, Bogdan_2024, juodzbalis_2024, Maiolino_2024}, and perhaps even shedding light on the seeding mechanisms of SMBHs. Notably, \textit{JWST} has identified dual AGN at rates an order of magnitude higher than predictions from cosmological simulations~\citep{perna2023ga, perna2023surprisingly, Ubler_2024}, complementing earlier detections made using other electromagnetic (EM) instruments~\citep{Rodriguez_2006, Comerford_2013, Goulding_2023}, and upcoming observations, by for example Euclid~\citep{euclid_2025}. However, resolving the two SMBHs involved in a merger at small separations remains a significant challenge, with only a handful of resolved systems being confirmed~\citep[see e.g.][]{Rodriguez_2006, trindade_2024}. This limits the insights that can be gained into SMBH mergers through EM observations alone. 

The advent of GW astronomy, marked by the groundbreaking detection of merging stellar-mass BHs by LIGO~\citep{abbott_2016}, has revolutionized our ability to study such phenomena. Advanced LIGO is now pushing into the intermediate mass BH regime, with a binary merger of mass $\sim 150$~\Msun{} being detected~\citep{abbott_2020}. Pulsar timing arrays (PTAs) have already provided evidence for a stochastic GW background (SGWB) in the nHz range, likely sourced by a population of merging SMBHs at the high-mass end~\citep{agazie_2023, Agazie_2023b, antoniadis_2023, Reardon_2023, Xu_2023}. In the next 10 -- 15 years, the Laser Interferometer Space Antenna (\textit{LISA}) will be capable of detecting SMBH mergers in the mass range of $10^4 - 10^7$ \Msun{} out to redshifts as high as $z \sim 20$~\citep{klein_2016, amaroseoane_2017, Katz_2018}. These GW observations will provide insights into SMBH merger dynamics and probe the SMBH population~\citep{agazie_2023, sato_2023, epta_2024}. Therefore, it is crucial to develop accurate theoretical models of SMBH mergers to ensure that GW and EM data from these instruments can be interpreted effectively.

Cosmological hydrodynamical simulations have been essential in understanding the population of SMBHs in our Universe and their relation with host galaxy properties~\citep[see e.g.][]{sijacki_2015, Habouzit_2021}. In anticipation of upcoming GW observations, these state-of-the-art simulations are now being used to better understand the merging population of SMBHs and make predictions for both merger rates~\citep{Kelley_2016, Khan_2016, Salcido_2016, Chen_2022, kunyang_2024, chen_2025} and complementary EM observations of dual AGN~\citep{rosas_2019, volonteri_2022, chen_2023, puerto_2025}.

Despite their varying degrees of success in explaining the properties of SMBHs, a fundamental challenge for these simulations lies in the incomplete picture of SMBH seeding mechanisms~\citep{Inayoshi_2020}. \Fable{}, the simulation used in this work~\citep{Henden_2018, Henden_2019, Henden_2019b}, together with other widely used simulations adopt a simplistic seeding prescription, placing BH particles with an initial mass of $\sim 10^5$~\Msun{} at the centres of haloes that exceed a certain threshold mass~\citep{Sijacki_2007, Di_Matteo_2008, Henden_2018}. Although recent efforts have aimed to improve these seeding models~\citep{DeGraf_2019, bhowmick_2024}, they remain a major limitation in capturing the true nature of the formation of SMBHs and can significantly affect not just the SMBH number density but also SMBH population mass growth through the merging channel.

SMBH mergers in cosmological simulations are also hindered by numerical effects. These arise from the coarse assumptions necessary to make simulations of large cosmological volumes computationally feasible. A primary limitation is the relatively low spatial resolution in large-scale cosmological simulations, which typically causes BHs to merge at separations of several kiloparsecs, bypassing many of the important hardening mechanisms discussed earlier~\citep{vogelsberger_2013}. This issue is well recognised, and to address it, many authors have applied post-processing techniques to estimate time delays at sub-kpc scales, based on the properties of the host galaxy of the remnant BH~\citep{Kelley_2016, Kelley_2017, sayeb_2020, Li_2022}. Idealised simulations, performed at much higher spatial resolution, have also been employed to study SMBH mergers in greater detail and better constrain hardening processes at specific scales~\citep{siwek_2023, bourne_2024}, although this comes at the cost of sacrificing the statistical power derived from larger volumes. More recently, significant progress has been made in incorporating these hardening mechanisms directly into cosmological simulations~\citep{Mannerkoski_2021, kunyang_2024}. In addition to missing out on these important hardening mechanisms, a consequence of low spatial resolution is that BH particles might be merged prematurely when two host galaxies undergo close encounters during the galaxy merger process~\citep{Damiano_2024}. 

Another significant limitation is the challenge of modelling DF acting on BH particles in cosmological simulations. In these simulations, the masses of surrounding DM, stellar, and gas resolution elements are often comparable to or exceed those of the BH particles, especially before the BHs have been able to significantly gain mass, leading to erratic scattering by the environment and the consequent `heating' of the BH orbit~\citep[see e.g.][for a general discussion and analysis of these issues]{Power_2003, Ludlow_2019}. This issue is closely related to, and further exacerbated by, the need to `soften' gravitational interactions. In typical cosmological simulations, the relatively large softening length value (of the order of $\sim 10^2-10^3$~pc), by definition reduces the gravitational interaction strength between the BH and surrounding matter below this scale. This effectively sets a lower limit on the minimum impact parameter that contributes to the drag force on the BH thus greatly limiting our ability to resolve DF in cosmological simulations~\citep{Merritt_1995,athanassoula_2000,Power_2003, genina_2024}. As a result, SMBH orbits in simulations are frequently unrealistic, with BHs either wandering away from the central regions of their host galaxies or failing to sink efficiently to the galactic centre. Since galactic nuclei are typically the densest regions of galaxies, these inaccuracies in BH orbits can lead to erroneous SMBH accretion rates, affecting the strength of feedback processes and, as a consequence, the environmental coupling of SMBHs~\citep{Callegari_2011, ragone_2018}. Moreover, the motion of BH particles involved in numerical SMBH mergers, even when above the spatial resolution, can remain unrealistic.

Orbit heating and wandering of BH particles are well-known numerical artefacts in cosmological simulations, and different approaches have been developed to mitigate these effects. For example, the Illustris and IllustrisTNG simulations~\citep{genel_2014, vogelsberger_2014, sijacki_2015, weinberger_2016, Pillepich_2017}, and similarly \Fable{}~\citep{Henden_2018}, adopt a `repositioning' technique. In this method, BH particles are moved towards a local minimum of the gravitational potential within the BH smoothing length at each time step~\citep{springel_2005b, Di_Matteo_2008, sijacki_2015, ragone_2018, bahe_2022}. This approach ensures that BHs largely remain anchored to the central regions of their hosts, enabling more realistic accretion rates. However, it results in BH dynamics that are inherently unphysical.
More recent efforts have focused on incorporating unresolved DF directly into simulations using sub-resolution prescriptions~\citep{Petts_2015,Petts_2016, tremmel_2015,Pfister_2019, bird_2022, chen2022dynamical, Ma_2023, Damiano_2024}. These models aim to approximate DF effects more accurately but often need to rely on idealized assumptions, such as specific velocity distributions and uniform background densities, which are rarely valid in complex astrophysical environments like galaxy mergers. \citet{genina_2024} proposed a correction for unresolved DF using calibrations based on the KETJU code, which employs post-Newtonian corrections to avoid gravitational softening in interactions of BHs and surrounding stars~\citep{Rantala_2017, Mannerkoski_2023}. While all these methods offer unique advantages, they also have limitations. Notably, none of these approaches can accurately model DF for BH particles with masses comparable to those of DM particles, which, with present mass resolution requirements of cosmological simulations, correspond to the crucial BH seed growth and merger regime. As a result, these methods are either not yet robust enough and/or too expensive for widespread implementation in large-scale cosmological simulations, where accurately capturing BH dynamics across a range of environments is essential.

On the other hand, DF on galactic scales acting on galaxies is well-modelled in cosmological simulations since it depends largely on `macrophysical' properties of the galaxies, as opposed to the `microphysical' properties of the BH particle and its surrounding which dictate the motion of SMBHs under DF~\citep{boylan_2007, lotz_2008}. In this paper, we leverage this success to explore how numerical limitations, such as low spatial resolution and unresolved DF on BHs, influence BH merger rates in the \Fable{} simulation. Our focus is on the initial stage of BH hardening — namely, DF operating \textit{above} the resolution limit of the simulation. We address this in post-processing by tracking the merger process of the host galaxies within the simulation and delaying the BH merger event until the host galaxies have merged. We emphasize that these time delays arise during the well-resolved phase of galactic dynamics, and are in addition to those caused by sub-resolution hardening mechanisms. During these time delays, we also allow the BHs to accrete based on the properties of their host galaxies. We then evaluate the implications of these time delays, together with the prolonged accretion they result in for future EM and GW observations.

The rest of the paper is organised as follows. In Section~\ref{sec:bh-population} we describe the \Fable{} simulation used for this work, and validate the simulated BH population against key observational datasets. In Section~\ref{sec:mergers-fable} we turn our attention to the SMBH merger population in \Fable{}, investigating the dynamics of BHs during galaxy interactions, including the numerical effects leading to premature mergers, and the observational consequences. We discuss our main findings and conclude in Section~\ref{sec:conclusion}.

\section{The BH population in FABLE} \label{sec:bh-population}

In this work, we use the simulation suite \Fable{} to investigate the robustness of SMBH merger rates such simulations can predict. In this section, we will first start by analysing the population of SMBHs in \Fable{}. This is not only crucial to verify the validity of using these simulations to estimate the population of SMBH mergers, but is also necessary when interpreting our results in the context of present and future observations. The \Fable{} simulation suite is extensively described by \citet{Henden_2018, Henden_2019, Henden_2019b} and its extension to a larger cosmological volume by \citet{bigwood_2025}. A brief summary is given here, with an emphasis on the modelling of BH particles in the simulation.

\subsection{Methodology: the \Fable{} simulation} \label{sec:methods-fable}

\subsubsection{Basic simulation properties} \label{sec:fable_basics}
The \Fable{} simulations employ the massively parallel code \textsc{Arepo} which solves the hydrodynamic equations on a Voronoi mesh in a quasi-Lagrangian fashion~\citep{Springel_2010, Pakmor_2015}. Gravitational interactions are modelled using the TreePM approach whereby stars, DM and BHs are represented by collisionless particles. 

\Fable{} employs a variety of sub-grid models that simulate processes important for galaxy formation. These models are largely based on those employed in the Illustris simulation~\citep{vogelsberger_2013, vogelsberger_2014, torrey_2014, sijacki_2015}, including descriptions of star formation~\citep{springel_2003}, radiative cooling~\citep{katz_1996, Wiersma_2009}, and chemical enrichment~\citep[based on work by][]{Wiersma_2009b}. To improve agreement with observations, the models describing stellar feedback~\citep{vogelsberger_2013} and AGN feedback~\citep{dimatteo_2005, Sijacki_2007} were updated in \Fable{} and calibrated against the local galaxy stellar mass function and the gas mass fractions of observed massive groups and clusters~\citep{Henden_2018}.

The \Fable{} simulation suite we use in this work consists of two large-volume cosmological simulations, with periodic lengths of  $40 \, h^{-1}$~cMpc and $100 \,h^{-1}$~cMpc. We will refer to these as the smaller and larger \Fable{} boxes respectively when distinguishing between the two. Since we are interested in the cosmological population of SMBH mergers, including very massive BHs relevant for the PTA regime, we primarily use the larger box in this work, hereafter simply referred to as \Fable{}. 

This cosmological box is evolved from initial conditions based on a Planck cosmology~\citep{planck_2016} with cosmological parameters $\Omega_\Lambda = 0.6935$, $\Omega_M = 0.3065$, $\Omega_b = 0.0483$, $\sigma_8 = 0.8154$, $n_s = 0.9681$ and $H_0 = 67.9$~km s$^{-1}$~Mpc$^{-1} = h\times 100$~km s$^{-1}$~Mpc$^{-1}$. The box contains $1280^3$ DM particles with mass $m_{\rm DM} = 3.4\times10^7\,h^{-1}$~\Msun{} and an initial number of $1280^3$ gas resolution elements with a target mass of $\overline{m}_b = 6.4\times 10 ^6 \, h^{-1}$~\Msun{}, from which stars and BHs form. The gravitational softening length is set to $2.393 \, h^{-1}$~kpc in physical coordinates below $z = 5$ and in comoving coordinates for higher redshifts. The simulation is evolved until $z=0$, with the internal parameters of all the particles and gas cells in the simulation being stored in 136 snapshots.

\subsubsection{Modelling the BH population} \label{sec:Fable_BHs}

BHs are modelled as massive, collisionless sink particles in \Fable{}. These particles are seeded with a mass of $10^5 \,h^{-1}$~\Msun{} in DM haloes with a mass exceeding $5\times 10^{10} \,h^{-1}$ \Msun{} that do not already host a BH particle. During seeding, the highest density gas cell in the halo is converted into a BH particle~\citep{Sijacki_2007, Henden_2018}. The smoothing length of a BH particle is defined as the comoving radius of the sphere enclosing the 64 nearest-neighbour gas cells around the BH~\citep{Nelson_2019}. 

As in Illustris, and many other large-scale cosmological simulations~\citep{vogelsberger_2014, Crain_2015, Pillepich_2017,Schaye_2023}, \Fable{} uses a repositioning scheme to `anchor' the BH particles to the minimum of the host halo's gravitational potential. At each time-step, the simulation identifies the particle with the lowest gravitational potential within a sphere of radius equal to the BH's smoothing length, and relocates the BH particle to that position. This prevents spurious BH movement caused by two-body scattering with DM or star particles which have masses comparable to that of the BH particle, and is intended to ensure that BHs reside at the centres of host galaxies to allow BHs to accrete and grow in a realistic way~\citep{Sijacki_2007, sijacki_2015, bahe_2022}. 

Throughout the simulation, these BH particles grow via mergers and accretion~\citep{Sijacki_2007, vogelsberger_2013, sijacki_2015}. Accretion happens at an Eddington-limited Bondi-Hoyle-Lyttleton-like rate
\begin{equation} \label{eqn:accretion}
    \dot{M}_{\rm BH} = \min\left[{\frac{4 \pi \alpha G^2 M_{\rm BH}^2 \rho}{(c_s^2 + v_{\rm BH}^2)^\frac{3}{2}}}, \dot{M}_{\text{Edd}}\right]\,,
\end{equation}
where $\rho$ and $c_s$ are the density and sound speed, respectively, of the surrounding gas, $v_{\rm BH}$ is the relative velocity of the BH with respect to the surrounding gas, and $\alpha = 100$ is the boosting factor \citep[see e.g.][for a discussion about $\alpha$]{sijacki_2009}. As a consequence of the repositioning scheme described above, BH velocities in the simulation are not physically meaningful and do not represent the movement of a SMBH in a realistic astrophysical environment. Because of this, the relative velocity is neglected when calculating the accretion rate using equation~\eqref{eqn:accretion}, which clearly may lead to overestimation of the accretion rates~\citep{vogelsberger_2013}. The Eddington accretion rate is given by
\begin{equation} \label{eqn:eddington}
    \dot{M}_\text{Edd} = \frac{4\pi G M_\text{BH} m_p}{\epsilon_r c \sigma_T}\,,
\end{equation}
where $m_p$ is the proton mass and $\sigma_T$ is the Thomson cross-section.
Accretion happens assuming a radiative efficiency of $\epsilon_r = 0.1$~\citep{Yu_2002, springel_2005}; a fraction $(1-\epsilon_r)$ of this accreted mass is added to the internal mass of the BH particle, while the rest is made available as feedback energy. 

In certain cases, the BH particle is located in an environment with insufficient gas pressure to compress the gas and raise the density higher than the star formation threshold. In such cases, the $\alpha$ pre-factor that is used to account for the unresolved hot and cold phases of the sub-grid ISM could lead to the formation of an unphysically large and hot gas bubble around the BH accreting from a low-density medium. This leads to unrealistically high accretion rates. To correct for this, in Illustris and \Fable{} the accretion rate is multiplied by a pre-factor $(P_{\rm ext}/P_{\rm ref})^2$, when the pressure in the environment of the BH particle $P_{\rm ext}$ is less than $P_{\rm ref}$, a reference pressure calculated by balancing the cooling losses in the bubble with the injected feedback energy from the BH. For further details on this pressure criterion see \citet{vogelsberger_2013}.

Feedback operates in one of two modes, and is determined by the Eddington fraction of the BH, $f_\text{Edd} = \dot{M}_\text{BH}/\dot{M}_\text{Edd}$. For $f_\text{Edd} > \chi_\text{radio}$, a threshold set to 0.01 in the \Fable{} model, feedback occurs in quasar mode in which a fraction of the bolometric luminosity ($\epsilon_f = 0.1$) is coupled thermally to the surrounding gas and energy transfer happens isotropically~\citep{dimatteo_2005, springel_2005b}. This commonly occurs at high redshift when cool gas is abundant and drives efficient BH accretion and for lower BH masses for which it is easier to reach high Eddington fractions. For $f_\text{Edd} < \chi_\text{radio}$, feedback happens in radio mode in which hot bubbles are injected into the surrounding gas at random spatial positions to mimic lobe inflation by an unresolved AGN jet~\citep{Sijacki_2007}. For further details on the radio and quasar modes in \Fable{} see \citet{Henden_2018, bigwood_2025}.

Furthermore, in Section~\ref{sec:fable-results} we study the quasar luminosity function (QLF) resulting from the BH population in \Fable{}. In order to calculate the bolometric luminosity of our BH particles due to accretion, we first assume that all BHs are radiatively efficient with luminosities given by
\begin{equation}
    L_\text{bol} = \epsilon_r \dot{M}_\text{BH} c^2\,.
\end{equation}
We also study the impact of a separate treatment which accounts for the radiatively inefficient AGN population, i.e.~BHs accreting at low Eddington fractions. For these BHs, we use the approach outlined by \citet{Koudmani_2024}. This uses results from an analytic fitting function by \citet{Xie_2012} corrected using general relativity and radiative transfer simulations~\citep{Ryan_2017}. Based on $f_\text{Edd}$ the radiative efficiency $\epsilon_r$ is then given by
\begin{equation} \label{eqn:radiative}
    \log \epsilon_r =
    \begin{cases} 
    -1.0 &\text{for } f_\text{Edd} \geq 0.1 \, \\[10pt]
    -1.0 - \left(\frac{0.0162}{f_\text{Edd}}\right)^4 & \text{for } 0.023 < f_\text{Edd} < 0.1\,, \\[10pt]
    \sum_n a_n (\log f_\text{Edd})^n & \text{for } 10^{-4} < f_\text{Edd} \leq 0.023\,, \\[10pt]
    \sum_n b_n (\log f_\text{Edd})^n & \text{for } f_\text{Edd} \leq 10^{-4}\,.
    \end{cases}
\end{equation}
The fitted values are $a_0 = -0.807$, $a_1 = 0.27$, $a_n = 0 \, \,(n \geq 2)$, $b_0 = -1.749$, $b_1 = -0.267$, $b_2=-0.07492$ and $b_n =0 \,\,(n \geq 3)$~\citep{inayoshi_2019}.

\subsubsection{BH mergers} \label{sec:bh_mergers_fable}

Two BH particles are considered to be merged in \Fable{} when they come within a particle smoothing length of each other. This is an adaptive length that is used to estimate hydrodynamic properties and is usually on the order of a kpc. Because of the repositioning technique used, a BH merger happens irrespective of the relative velocities of the two BHs~\citep{sijacki_2015}. A merger happens instantaneously, with the remnant BH having a mass equal to the sum of the masses of the merging BHs. To avoid confusion, we will refer to a merger event in the \Fable{} simulation as a `numerical BH merger' from now on. 

In addition to the information stored in the simulation snapshots mentioned in Section~\ref{sec:fable_basics}, BH specific output is also recorded at every time step, giving better time resolution for BH accretion rates, local densities around the BHs and numerical merger events. For each merger event, the scale factor and the masses of both BHs are recorded at the exact time of the merger. After making some selection cuts, which are discussed in Section~\ref{sec:selection-cuts}, this merger catalogue is used in our analysis. Note that this is the first work to use \Fable{}, with its more realistic galaxy groups and clusters, to study this merging population of BHs.

\subsubsection{Host galaxy identification} \label{sec:host-galaxy}

To identify gravitationally bound structures on-the-fly in the simulation, \Fable{} uses the friends-of-friends (FoF) and \textsc{Subfind} algorithms~\citep{davis_1985, springel_2001, dolag_2009}. The FoF algorithm first identifies `haloes' by grouping together DM, stars and gas particles using a characteristic linking length. The \textsc{Subfind} algorithm then splits these haloes into self-bound substructures called `subhaloes' which can roughly be interpreted as central and satellite galaxies surrounded by their gaseous and DM haloes. This is done in configuration-space by identifying density peaks. The `central subhalo' is the subhalo at the minimum gravitational potential of the FoF halo, while the other subhaloes in the FoF halo are considered to be satellites.

BH particles, like other particles in the simulation, can also be associated to haloes and subhaloes. According to the BH seeding prescription used in \Fable{} (see Section~\ref{sec:Fable_BHs}), BHs are seeded in central subhaloes. However, their host subhaloes can change via a variety of physical processes, including dynamical interactions with other massive structures, or BH mergers themselves. Subhaloes might also host multiple BH particles because of these interactions. In order to study the connection between host galaxies and their central BHs, we choose the most massive BH particle associated to a subhalo.

To track subhaloes across snaps, subhalo merger trees were constructed using the \textsc{Sublink} algorithm~\citep{Rodriguez_Gomez_2015}. In this algorithm, progenitors and descendents of subhaloes are identified based on their number of shared particles and their binding energies. In the context of this work, these subhalo merger trees are crucial in tracking the host subhaloes of BHs involved in a BH merger event across cosmic time, as will be explained in more detail in Section~\ref{sec:premature}.

\subsection{Results: Validating the BH population in \Fable{}} \label{sec:fable-results}
Throughout this section, we will present the results for the BH population in the \Fable{} box with a side length of $100\,h^{-1}$~cMpc, which is then used for BH merger predictions in Section~\ref{sec:mergers-fable}, and compare this to the results from the \Fable{} box with side length $40 \, h^{-1}$~cMpc presented by \citet{Henden_2018, Koudmani_2021}. When available, we will also draw comparisons to observational data.

\begin{figure*}
  \centering
  \begin{subfigure}[b]{\columnwidth}
    \centering
    \includegraphics[width=\columnwidth]{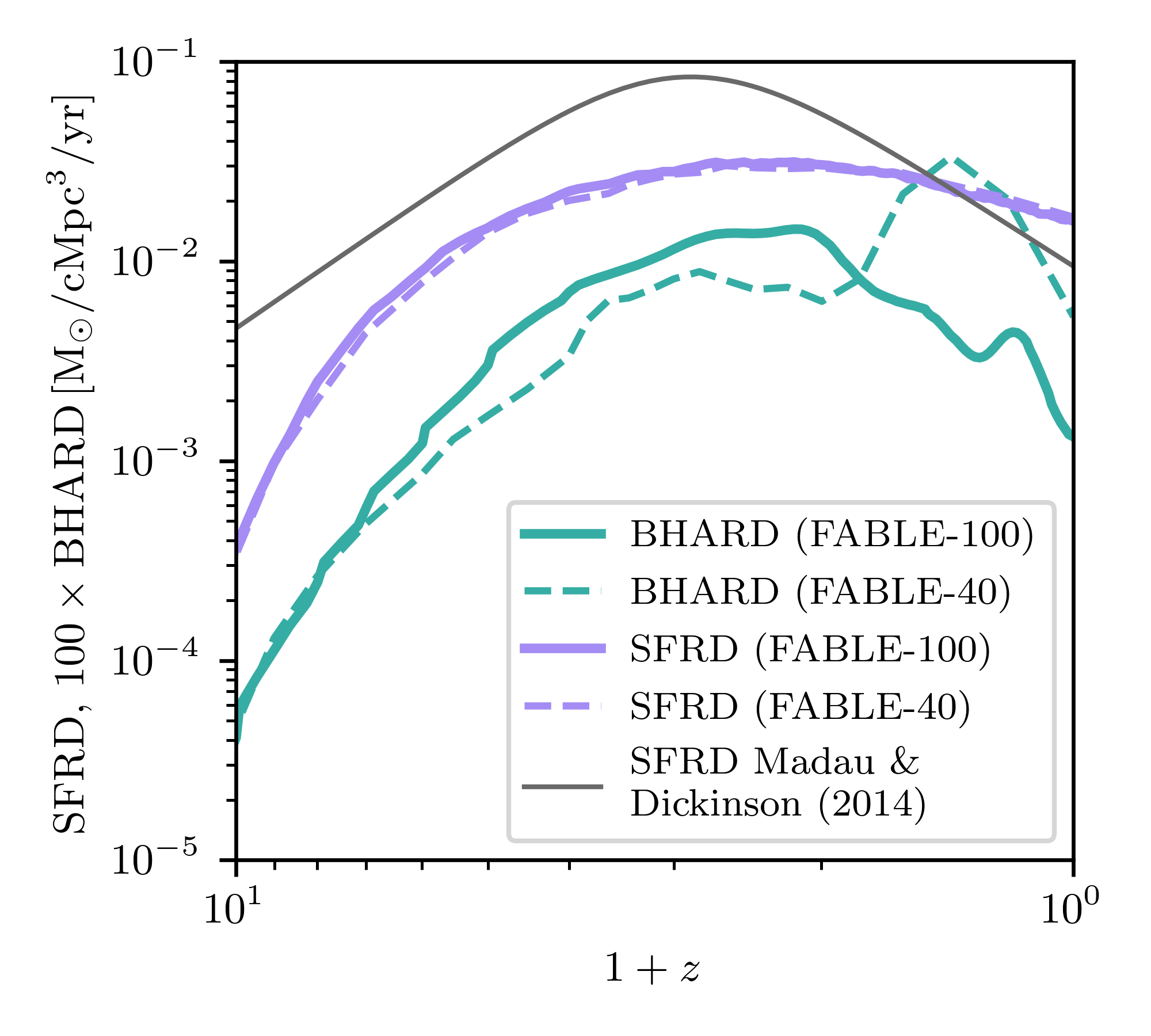}
    \label{fig:sfrd_bhard}
  \end{subfigure}
  \hfill
  \begin{subfigure}[b]{\columnwidth}
    \centering
    \includegraphics[width=\columnwidth]{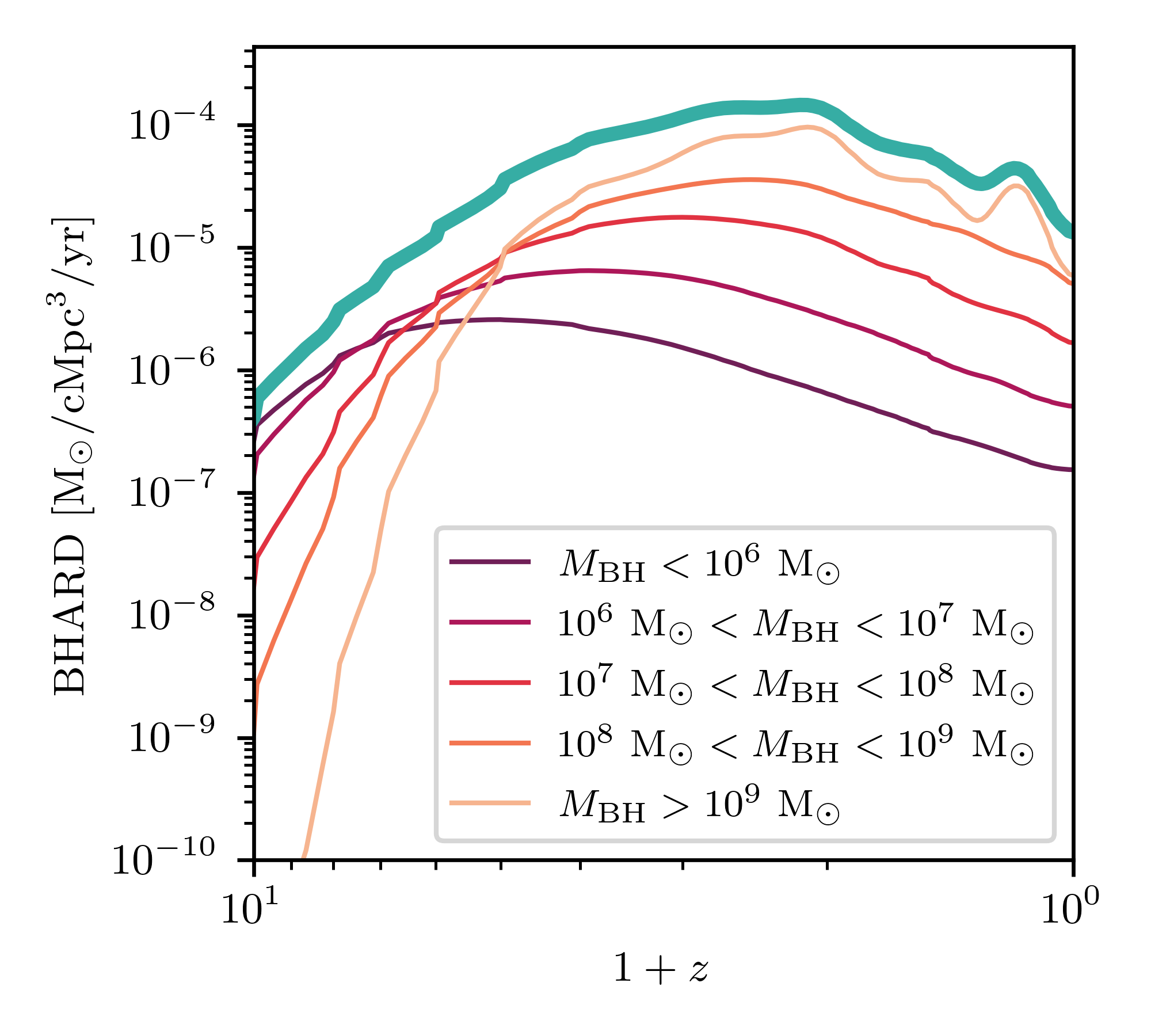}
    \label{fig:smd_bhmd}
  \end{subfigure}
  \caption{Left: The SFRD (purple curves) and the BHARD, scaled up by a factor of a 100 (teal curves), as a function of redshift for the \Fable{} simulation. The dashed lines shows results for the \Fable{} box with side length of $40\, h^{-1}$~cMpc (\Fable{}-40), while the solid lines correspond to the box with side length of $100 \, h^{-1}$~cMpc (\Fable{}-100). The \Fable{} accretion rate densities are smoothed using a 1D-Gaussian filter (see main text for details). In grey we show observations for the SFRD from \citet{Madau_2014}. Right: the total BHARD (thick teal line) and the BHARD split according to BH mass (coloured lines according to the legend) as a function of redshift for the larger \Fable{} box. The BHARD lines are again smoothed using a 1D Gaussian filter. }
  \label{fig:bhard}
\end{figure*}

The left-hand plot of Fig.~\ref{fig:bhard} shows the star formation rate density (SFRD) and the BH accretion rate density (BHARD). The SFRD from \Fable{} is calculated by summing the total SFRD within the stellar half-mass radius of all subhaloes at a specific redshift. This is then compared to the best fit from \citet{Madau_2014} which uses compiled data from UV and IR observations to obtain a comprehensive  model of cosmic star formation history. Note that for a direct comparison, we multiply their fit by a factor of 0.63 to convert from a \citet{salpeter_1955} IMF to a \citet{Chabrier_2003} IMF which is what is used in \Fable{}. Compared to the \citet{Madau_2014} fit, our SFRD is lower than observations at $z \gtrsim 1$, while being slightly higher than observed values at lower redshifts. The observed SFRD peaks at $z\sim 2$ while in \Fable{} the SFRD has a broad `high' plateau around this redshift. 

As discussed in Section~\ref{sec:Fable_BHs}, within the Bondi formalism, the accretion rate depends on the density and local sound speed of the gas surrounding the BH (see equation~\eqref{eqn:accretion}). Prevalently, `radio' mode AGN feedback causes fluctuations in the local density and temperature close to the accreting SMBHs which in turn cause the accretion rate to fluctuate. These fluctuations are smoothed using a 1D Gaussian filter in the BHARD curve so that the overall trend can be studied\footnote{The standard deviation of the Gaussian kernel is set to the time interval between four/one simulation snapshots for the larger/smaller \Fable{} box. The spacing between snapshots is adaptive but is on average $\Delta a \sim 0.007$ and $\Delta a \sim 0.04$ for the larger and smaller box, respectively.}. The differences in BHARD between the two \Fable{} boxes, which have the same mass and spatial resolution, is due to the differing box sizes as rarer objects corresponding to galaxy groups and clusters are less prevalent in smaller volumes. Thus the larger \Fable{} box will have a larger number of these rarer objects, leading to better population statistics. In the larger \Fable{} box, the peak in the BHARD (around $z\sim 1.2$) is more pronounced than the peak in the SFRD and is a factor of $\sim 200$ lower than the SFRD. The right-hand plot of Fig.~\ref{fig:bhard} shows the split of BHARD according to bins in BH mass for the larger \Fable{} box. The redshifts at which the BHARD for BHs at a particular mass peak reflect the cosmic time at which they contribute the most to the total BHARD with more massive BHs dominating at later times. This is due to a combination of their number density and accretion rate in absolute numbers.

\begin{figure*}
  \centering
  \begin{subfigure}[b]{\columnwidth}
    \centering
    \includegraphics[width=\columnwidth]{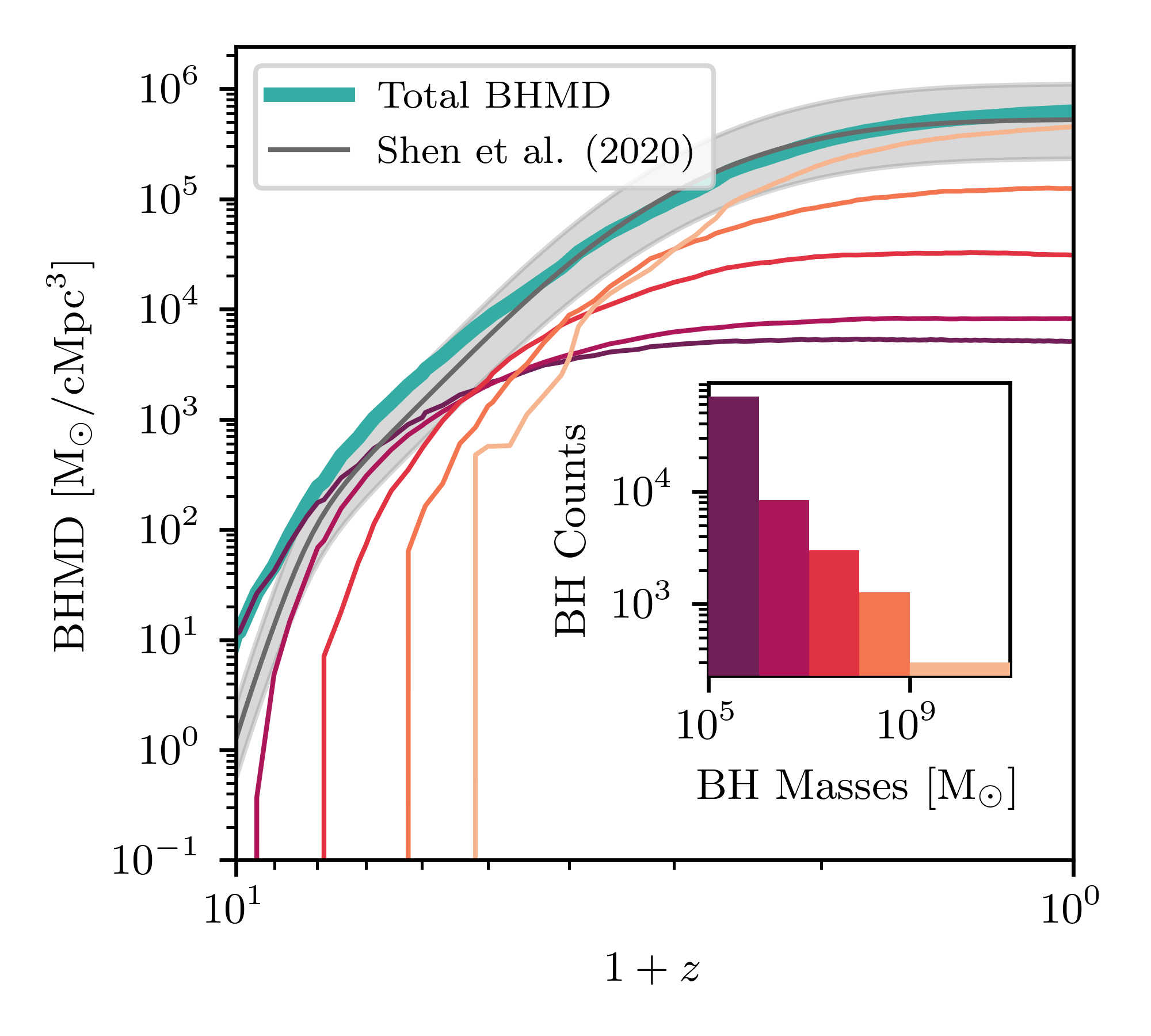}
    \label{fig:bhard_ranges}
  \end{subfigure}
  \hfill
  \begin{subfigure}[b]{\columnwidth}
    \centering
    \includegraphics[width=\columnwidth]{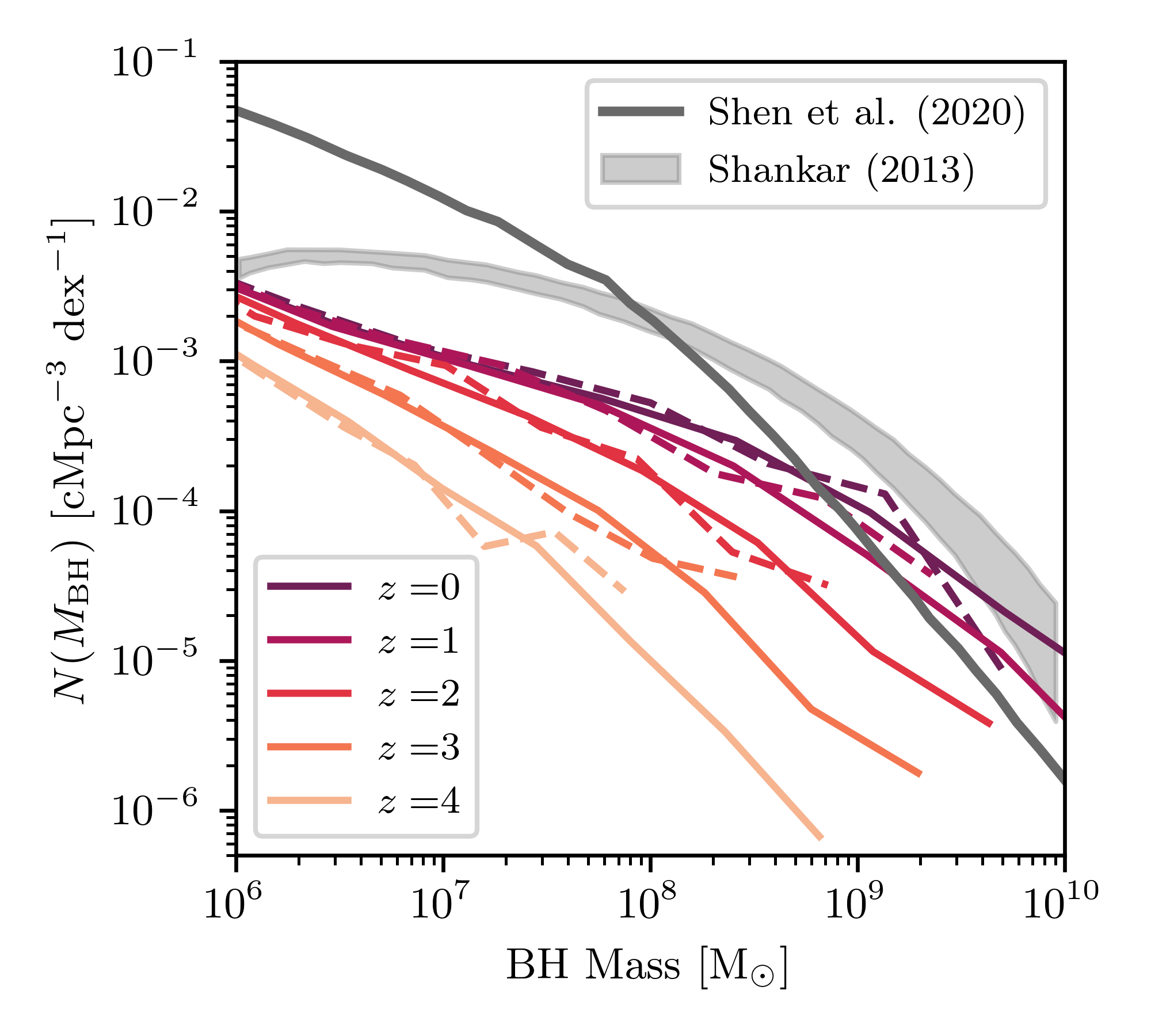}
    \label{fig:bhmd_ranges}
  \end{subfigure}
  \caption{Left: the BHMD as a function of redshift, split according to BH mass (coloured lines are as in the legend in the right-hand plot of Fig.~\ref{fig:bhard}), together with the total BHMD (thick teal line) for the \Fable{} box with side length of $100 \, h^{-1}$~cMpc. The thick grey line shows an estimate by \citet{Shen_2020} for the mass density of SMBHs derived from the bolometric luminosity function of AGN. This estimate assumes an average radiative efficiency $\epsilon_r = 0.1$ and a starting redshift of integration equal to 10. The grey shaded region shows the uncertainty in this estimate when increasing or decreasing $\epsilon_r$ by a factor of two. The inset plot shows the number of BHs in each mass bin at $z = 0$. Right: The BHMF at redshifts $z = 0, 1,2,3,4$, with colours as indicated by the legend, for all BHs in the \Fable{} simulation boxes. The solid lines correspond to the larger \Fable{} box, while the dashed lines correspond to the smaller box. The light grey shaded region shows the BHMF with $1\sigma$ uncertainty region calculated by \citet{Shankar_2013}, assuming the $M_{\rm BH}-\sigma$ relation defined by \citet{McConnell_2013} and applying it to all local galaxies. The thick grey line shows the BHMF calculated using the bolometric luminosity function of AGN in the local universe by \citet{Shen_2020} using their `deconvolution' method.}
  \label{fig:bh_masses}
\end{figure*}

The left-hand plot of Fig.~\ref{fig:bh_masses} shows the BH mass density (BHMD) split according to different BH mass bins as indicated in the legend of Fig.~\ref{fig:bhard}. As expected due to their growth via accretion (right-hand plot of Fig.~\ref{fig:bhard}) and mergers, more massive BHs dominate the total BHMD at later redshifts. The inset plot shows the total number of BHs in each mass bin at $z = 0$. Note that as a consequence of a limited simulation volume, the most massive BHs with mass $ > 10^9$~\Msun{} are scarce and hence results for this mass bin are tentative due to lack of statistics, although the larger \Fable{} box offers an improvement over the smaller box as discussed above. The total BHMD is shown in the thick teal line and is compared to the estimate by \citet{Shen_2020} obtained via the bolometric luminosity function of AGN observed out to $z \sim 6$ with a radiative efficiency set to $\epsilon_r = 0.1$ (solid grey line). We also show the uncertainty in this estimate when increasing or decreasing $\epsilon_r$ by a factor of two by the grey shaded region. The population of SMBHs in \Fable{} fits the observations very well, especially at lower redshifts.

The BH mass function (BHMF) at redshifts $z = 0, 1, 2, 3, 4$ is shown on the right-hand side of Fig.~\ref{fig:bh_masses}. The grey shaded region shows the mass function assuming the $M_{\rm BH} - \sigma$ relation by \citet{McConnell_2013} applied to all local galaxies with a $1\sigma$ uncertainty as calculated by \citet{Shankar_2013}. The thick grey line shows the BHMF obtained from the bolometric luminosity function of AGN in the local Universe using the `deconvolution' method discussed by \citet{Shen_2020}. \Fable{} underestimates the number density of BHs at all mass bins with respect to the estimates by \citet{Shankar_2013}. Despite the excellent agreement with the observed BHMD, \Fable{} is in disagreement with the constraint on the BHMF from \citet{Shen_2020}, with the number of low mass BHs being underestimated significantly \citep[see also detailed discussion on the lack of AGN in low mass galaxies by][]{Koudmani_2021}, and the number of high mass BHs being overestimated. This implies that although the \Fable{} simulation is likely correctly predicting the overall mass density of BHs, the distribution of this mass density across different mass bins is not quite in agreement with the latest observational estimates. This is driven by simplistic models of both BH seeding mechanisms and subsequent BH growth. The broader picture is similar across many widely-used cosmological simulations, none of which can reproduce the full range of observational constraints~\citep{Habouzit_2021}.

\begin{figure*}
    \centering
    \includegraphics[width = 2\columnwidth]{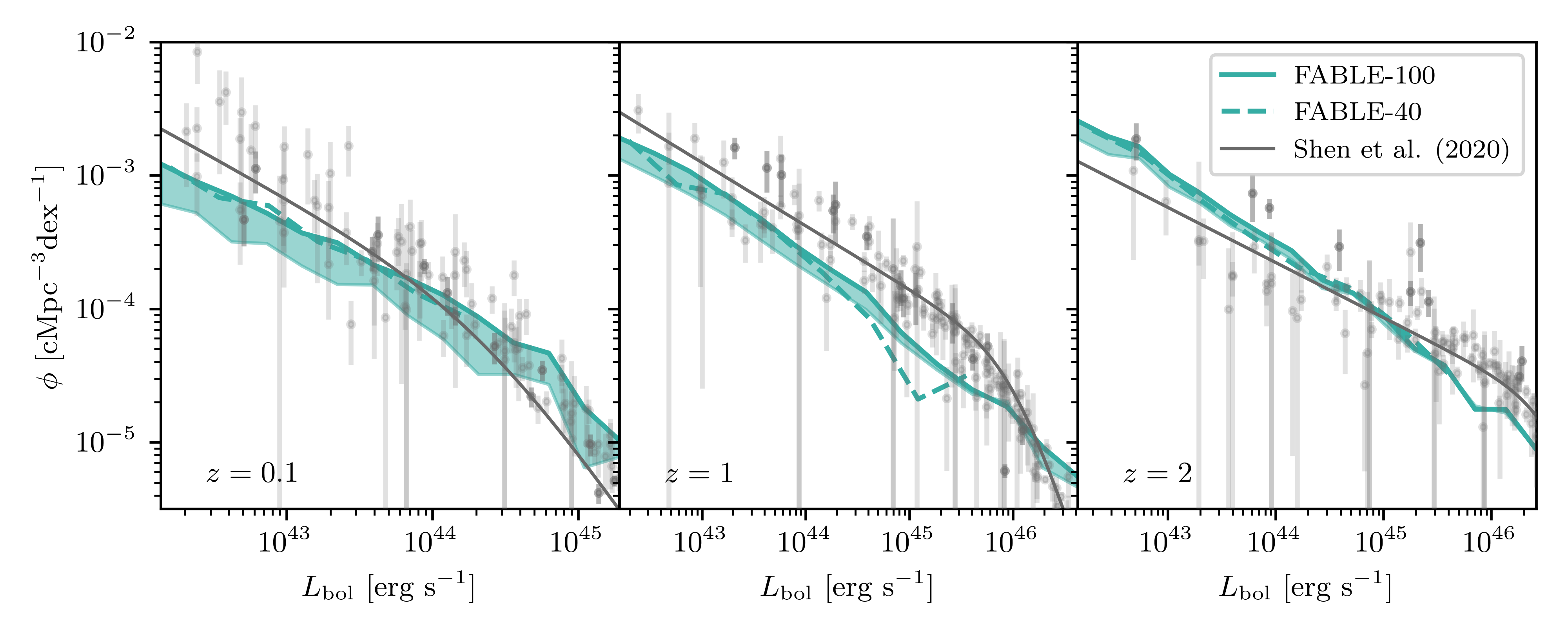}
    \caption{
    The AGN bolometric luminosity function at redshifts $z = 0.1, 1, 2$ from left to right.
    The solid teal lines show results from the larger \Fable{} box and are compared to the smaller box shown by the dashed teal lines. For the smaller \Fable{} box we assume for simplicity that all BHs in the simulation are radiatively efficient. For the larger \Fable{} box, we show the predicted luminosity function bracketed by this assumption and that accounting for radiatively inefficient AGN as discussed in Section~\ref{sec:Fable_BHs}. These simulation results are compared to the global fit by \citet{Shen_2020} made to observed AGN shown in the solid grey line (their global fit A is presented here). The grey points with error bars are the data used for this fit.} 
    \label{fig:bolometric}
\end{figure*}

With quasars being some of the most luminous objects in the Universe, it is crucial to check that the predicted QLF from \Fable{} matches observations. However, given that the bright quasars are scarce in our simulated volumes here we focus on the entire AGN population. Fig.~\ref{fig:bolometric} compares the AGN luminosity function of our two \Fable{} boxes with the global fit to observed AGN by \citet{Shen_2020} which compiles measurements made in the optical/UV, X-ray and infrared bands. We plot the results for $z=0.1$ (for easier comparison to observational data) and $z = 1,2$. For the simulations we first show the results under the assumption that all AGN are radiatively efficient. Additionally, for the larger \Fable{} box, we provide a bracketing of the predicted luminosity function by including a population of radiatively inefficient AGN, as described in Section~\ref{sec:Fable_BHs} and equation~\eqref{eqn:radiative}. Both \Fable{} boxes provide a good fit to the observed AGN luminosity function, especially within the context of results obtained from various cosmological simulations, as can be seen in Fig.~5 of \citet{Habouzit_2021}. Notably, \Fable{} avoids the over-prediction of the faint end of the luminosity function for higher redshifts which is seen in other simulations. As discussed by \citet{bigwood_2025} and \citet{koudmani_2022}, \Fable{} slightly underpredicts the QLF at the faint end at lower redshifts. This suggests that the observed SMBH population is likely accreting at more efficient rates than in \Fable{} in lower mass galaxies. 

\begin{figure}
    \centering
    \includegraphics[width=\columnwidth]{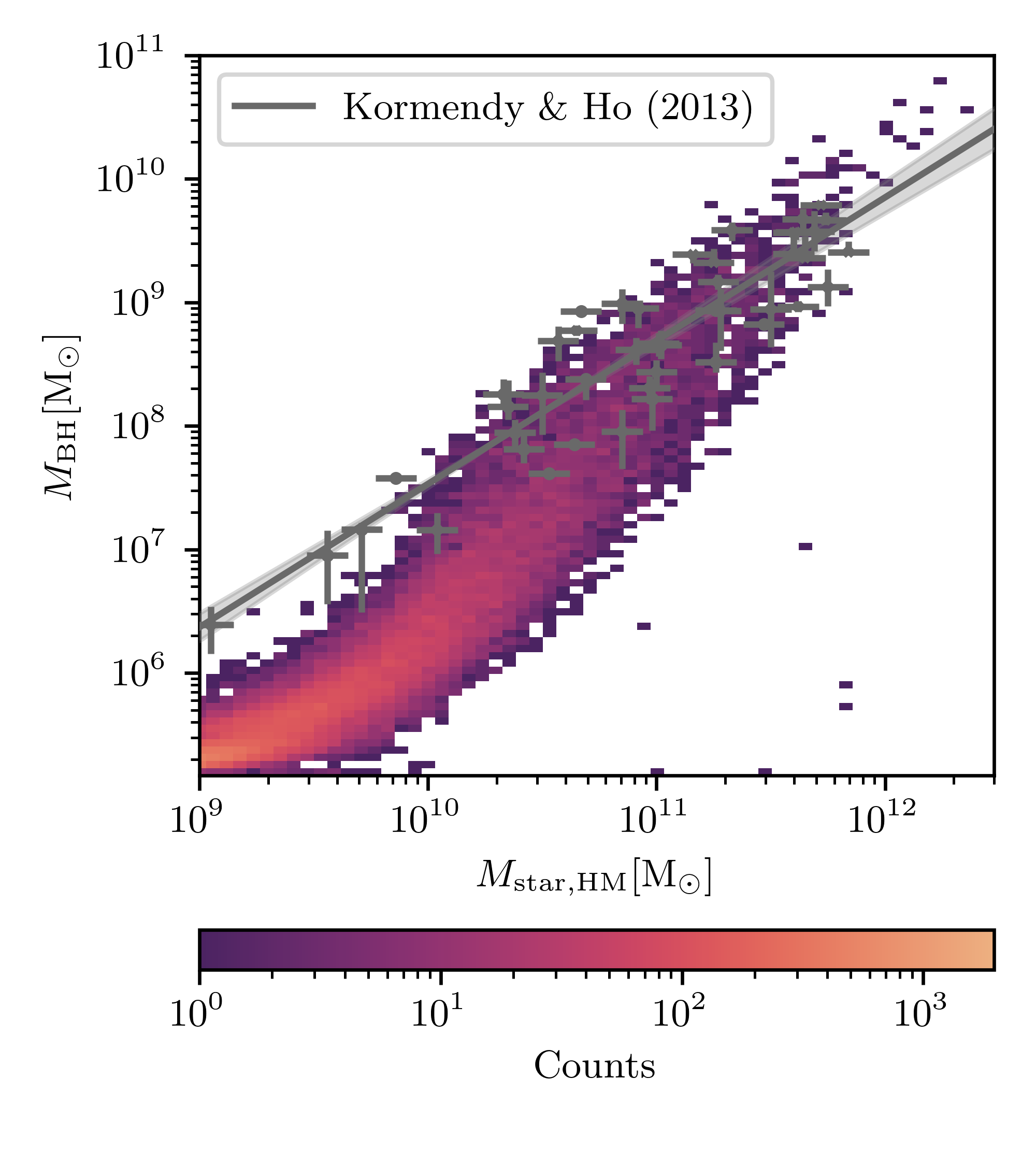}
    \caption{Central BH mass vs. stellar mass within the half-mass radius ($M_{\rm star, HM}$) of all galaxies (central subhaloes) in \Fable{} at $z=0$. The thick grey line shows the best fit for the local $M_{\text{BH}} - M_{\text{bulge}}$ relation by \citet{Kormendy_2013} fitted to ellipticals and galaxies with bulges only. The grey points with error bars are the data used for this fit. Notice that SMBHs in \Fable{} are less massive than predicted by the local relation at the low mass end, as also indicated by \citet{Koudmani_2021}.}
    \label{fig:mass-relation}
\end{figure}

Fig.~\ref{fig:mass-relation} shows the relationship between the mass of the central BH and the stellar mass within the half-mass radius, which is a simple estimate for the bulge mass, of the host subhalo. We note that using the stellar mass within the half-mass radius as a proxy for bulge mass is a common approach in cosmological simulations~\citep[see for example][]{Di_Matteo_2008, sijacki_2015} due to the challenges of morphological decomposition when comparing to observations~\citep{Scannapieco_2010, Dubois_2012}, though this simple approach might not capture the full complexity of bulge definitions. The best fit $M_\text{BH} - M_\text{bulge}$ relation by \citet{Kormendy_2013} is also shown together with data from spiral and elliptical galaxies used for this fit. While fundamental for understanding BH co-evolution with host galaxies, this mass relation has a large intrinsic scatter and becomes highly uncertain at the low mass end due to the difficulty in observing such low mass BHs. There is also a lack of consensus in the literature, with different authors finding somewhat different scaling relations~\citep{Kormendy_2013, McConnell_2013, reines_2015, greene_2020}. Keeping in mind that the \Fable{} simulation successfully reproduces the observed galaxy stellar mass function over a wide range of masses (see Fig. 1 by~\citet{bigwood_2025} and Fig. 2 by~\citet{Henden_2018}), Fig.~\ref{fig:mass-relation} shows that \Fable{} most likely underestimates the masses of SMBHs hosted by low (bulge) mass galaxies. This same trend is evident when considering the BH mass -- stellar mass relation (evaluated within twice the stellar half-mass radius), as also concluded by \citet{Koudmani_2021}, which is in line with our previous findings regarding the BHMF. Consequently, the SMBH merger rates we derive for BHs in the mass range $10^6\,\Msun{} \lesssim M_{\text{BH}} \lesssim 10^7\,\Msun{}$ should be interpreted as lower limits of the possible true rates. We investigate this issue further in Appendix~\ref{app:evolution}, where we plot BH mass as a function of the stellar mass within twice the stellar half-mass radius across different redshifts (Fig.~\ref{fig:relation_evolution}).

\section{The merging population of SMBHs in FABLE}
\label{sec:mergers-fable}

Building on the results in Section~\ref{sec:fable-results}, we now turn our focus to the merging SMBH population in the simulation. In this section we investigate how numerical effects lead to premature mergers within \Fable{} and present a first attempt at correcting for this in post-processing.

\subsection{Premature mergers due to numerical effects} \label{sec:premature}

A BH merger event in \Fable{} happens as a consequence of a subhalo interaction in which both subhaloes host a BH particle. In cosmological simulations, the merging process of subhaloes is often preceded by a series of close encounters, where the two subhaloes repeatedly approach and recede from one another before the final coalescence occurs. These interactions are driven by the dynamical interplay between the gravitational potential of the two subhaloes and the surrounding large-scale environment, as well as the DF exerted by the background matter. Such interactions can also occur during subhalo fly-bys without being necessarily succeeded by a subhalo merger. During these close encounters, it is common for the two BH particles to come within a smoothing length of each other, triggering a numerical BH merger event even though the final subhalo (i.e. galaxy) merger has not yet occurred. This effect is likely often of purely numerical nature; when the gravitational potentials of the host subhaloes are sufficiently well-defined for the \textsc{Subfind} algorithm to identify them as distinct structures, in many cases the BHs should remain physically bound to the centres of their respective subhaloes until the subhaloes themselves merge, if they do at all. Furthermore, recall that when two BHs are numerically merged, their relative velocities are not taken into account and there are no checks on whether they are gravitationally bound to each other. Ultimately, very high resolution simulations where the DF force is accurately followed are needed to settle this issue, but for the purpose of this work we assume that BHs should reside within the centres of their respective galaxies during the merging process as long as these galaxies can be reliably identified by the \textsc{Subfind} algorithm, which is a somewhat simplistic but much more reasonable physical scenario. 

In order to study the interplay between subhalo mergers and/or interactions with BH mergers in \Fable{}, we identify the host subhaloes of the two BHs involved in a numerical BH merger. We do this in the most recent snapshot before the numerical merger event where both BHs reside in the central subhalo of their respective FoF haloes. When there is no common snapshot in which both BHs are in central subhaloes, we identify the host subhaloes at the earliest snapshot in which both BHs are present in the simulation. 
We then track these host subhaloes using the subhalo merger trees constructed using the \textsc{Sublink} algorithm as discussed in Section~\ref{sec:host-galaxy}.

\begin{figure*}
    \centering
    \includegraphics[width = 2\columnwidth]{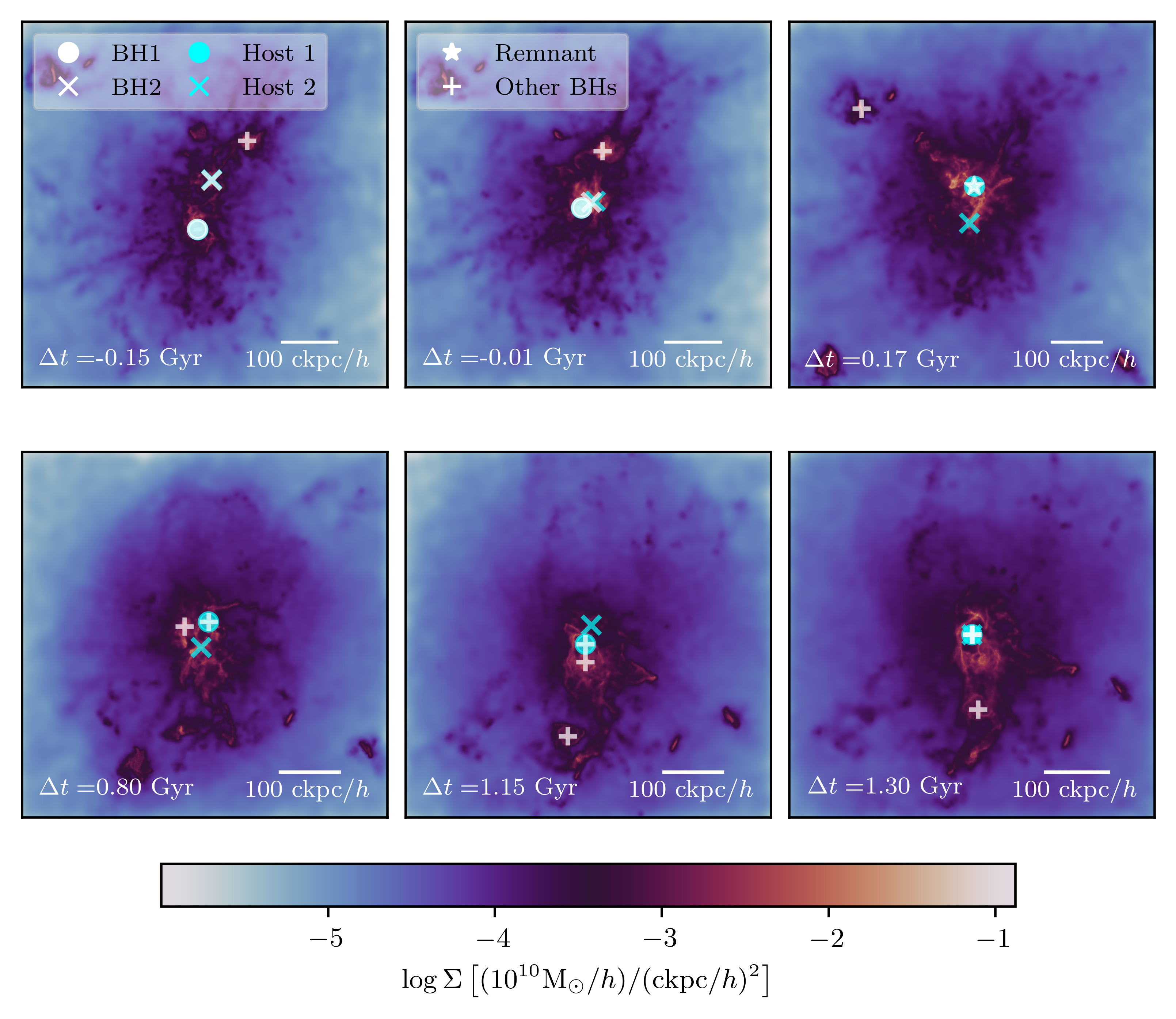}
    \caption{The projected gas density in a small cosmological volume surrounding a representative merger event in \Fable{} for six different snapshots in the simulation. The positions of the primary (BH1) and secondary (BH2) BHs are shown as a white circle and cross, respectively. The positions of their respective host subhaloes are shown using the same marker shapes but in cyan. The direct remnant of the BH merger of interest is shown as a white star, when present, while the positions of any other BHs in the vicinity are shown with a white plus symbol. The top-left panel shows the BHs two snapshots before the numerical BH merger, with both BHs residing at the centre of their respective host subhaloes. The top-centre panel corresponds to the snapshot directly preceding the numerical BH merger and shows BH2 moving away from the centre of its host subhalo due to the repositioning technique used in \Fable{}. The top-right panel shows the snapshot succeeding the merger, with the remnant BH residing in the host subhalo that hosted BH1 initially, and with the two subhaloes still resolved as two separate structures. The bottom-left panel shows the remnant BH merging with another BH particle (notice the change from a star marker to a plus marker), while on the bottom-right panel we see the two host subhaloes finally merging into one subhalo. $\Delta t$ in the bottom-left corner of each panel corresponds to the time elapsed since the numerical BH merger, with negative values indicating time to numerical BH merger. The scale in the bottom right corner of each plot is given in simulation units.}
    \label{fig:merger_process}
\end{figure*}

An example of a premature BH merger relative to the subhalo merger is illustrated in Fig.~\ref{fig:merger_process}. This figure shows the projected gas density in a small volume surrounding the merger event, highlighting the positions of the primary BH (BH1) and the secondary (BH2), their host subhaloes, and other nearby BHs in six different simulation snapshots. In the top-left panel, the two BHs are in the centres of their respective host subhaloes. At this cosmic time, the BHs have masses of $9.18 \times 10^7$~\Msun{} and $1.18\times10^7$~\Msun{} and reside in subhaloes with total stellar mass within twice the stellar half-mass radius of $2.66\times10^{11}$~\Msun{} and $6.39\times 10^{10}$~\Msun{}, respectively. The centre-top panel corresponds to the snapshot directly preceding the numerical BH merger, showing the two BHs still within their respective subhaloes, but with the secondary BH moving away from the centre of its host towards the strong gravitational potential of the primary's host. The top-right panel shows the snapshot directly succeeding the numerical BH merger, where the remnant BH is located at the centre of the subhalo that originally hosted the primary BH, while the BH2 host subhalo remains resolved as an independent structure. At this snap, the remnant BH has a mass of $1.16\times 10^8$~\Msun{}, with the stellar mass of its host galaxy being equal to $3.55\times10^{11}$~\Msun{}. The bottom-left panel shows the remnant BH undergoing a subsequent merger with another BH (the plus symbol indicates that this is no longer the direct remnant of the main merger event), and the bottom-right panel reveals the eventual coalescence of the two subhaloes into a single structure. This sequence unfolds over eleven snapshots for the selected merger event with an elapsed time of 1.3~Gyr between the numerical BH merger and the final coalescence of the two host subhaloes.

\begin{figure*}
    \centering
    \includegraphics[width = 2\columnwidth]{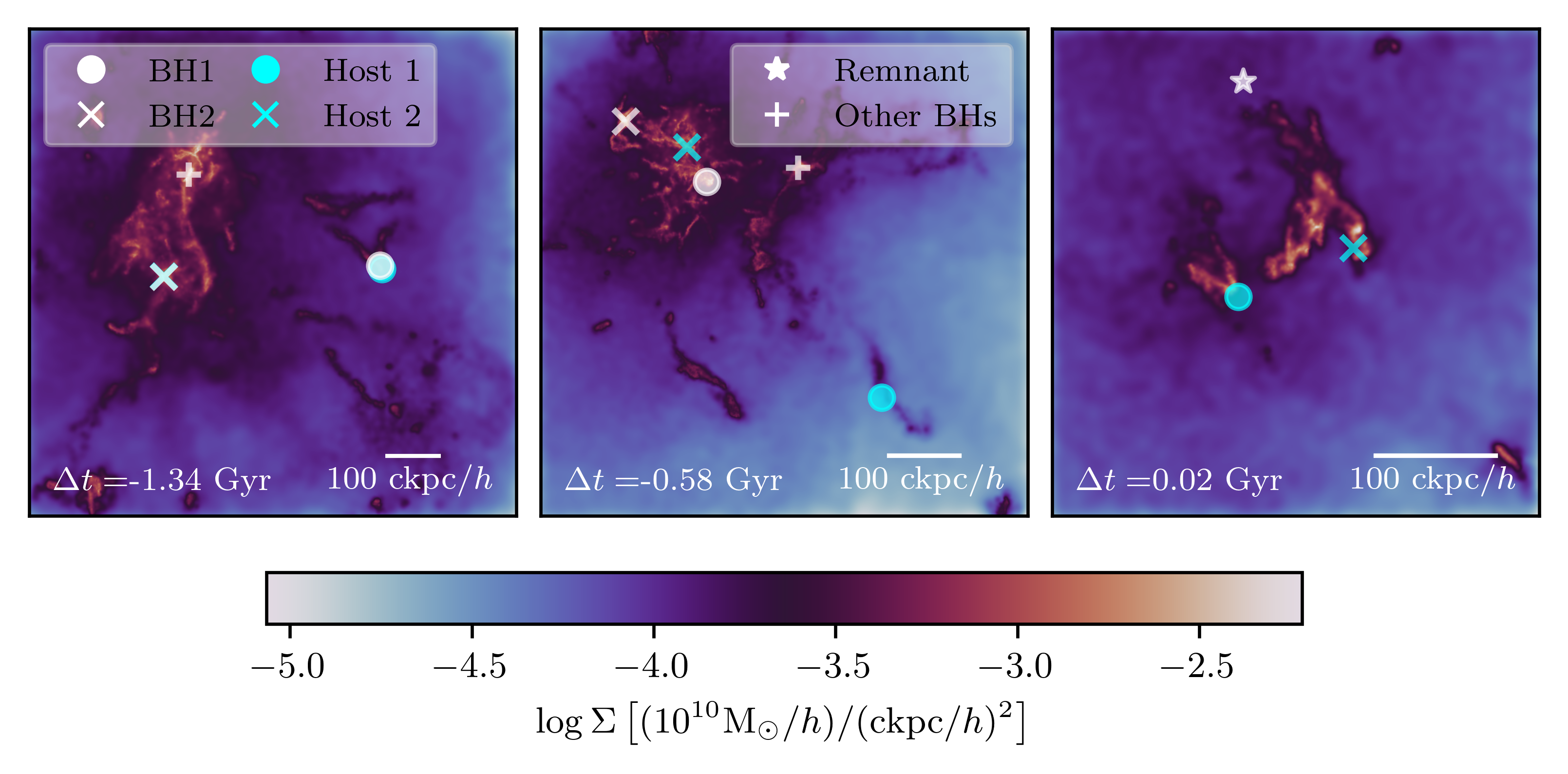}
    \caption{The projected gas density around a second representative BH merger event in \Fable{}. The legend is the same as described in the caption of Fig.~\ref{fig:merger_process}. The left panel shows the BHs residing at the centres of their respective host haloes. The centre panel shows the two BHs residing in completely different subhaloes than their original hosts while the right panel corresponds to the snapshot directly succeeding the numerical merger event, with the remnant BH residing in neither of the original host galaxies. The original host subhaloes do not merge by $z = 0$ in \Fable{}. }
    \label{fig:merger_process_2}
\end{figure*}

A separate but compounding numerical effect also occurs due to the repositioning scheme described in Section~\ref{sec:Fable_BHs}. During gravitational interactions with another subhalo, a BH particle can not only be displaced from the centre of its original host subhalo but, in some cases, also change its host subhalo entirely due to the strong gravitational influence of the interacting subhaloes. This effect is illustrated in Fig.~\ref{fig:merger_process_2} which shows a BH merger event where both BHs are removed from their original host subhaloes as a direct consequence of the repositioning scheme. At the snapshot where the host subhaloes are identified (left panel of Fig.~\ref{fig:merger_process_2}), the primary and secondary BHs have masses of $1.47\times 10^5$~\Msun{} and $1.60 \times 10^6$~\Msun{} respectively. They reside in subhaloes with total stellar mass in twice the stellar half-mass radius of $1.55 \times 10^{10}$~\Msun{} and $5.44\times 10^{10}$~\Msun{}, respectively. By the time of numerical BH merger, the primary BH grows to have a mass larger than the secondary. In the snapshot directly succeeding this numerical BH merger (right panel of Fig.~\ref{fig:merger_process_2}), the remnant BH has mass equal to $7.71 \times 10^8$~\Msun{} and resides in the central subhalo of the FoF halo with total mass equal to $1.01 \times 10^{12}$~\Msun{}. The original host subhaloes have masses of the order of $10^{10}$~\Msun{} and are substructures within the larger central subhalo whose strong gravitational potential completely displaces the remnant from the original satellite subhaloes hosting the BHs. 
Similarly to the premature mergers during subhalo interactions depicted in Fig.~\ref{fig:merger_process}, this displacement of BHs away from their original host subhaloes is purely a numerical artefact. Physically, BHs are expected to remain anchored at the centres of their host galaxies during interactions with passing galaxies, as long as the innermost dense region to which the BH is bound to exists, except in cases where those galaxies themselves merge. In some instances, the original host subhaloes of the merging BHs can go on to evolve independently, with a large spatial separation persisting between them. This can result in the original subhaloes even never merging at all, despite their BHs having merged within \Fable{}.

\subsection{Methodology: Macrophysical time delays to SMBH mergers} \label{sec:methods-delays}

In order to quantify the numerical effects described above, we introduce a time delay in post-processing corresponding to the interval between the numerical BH merger and the potential eventual merger of their host subhaloes. It is important to stress that these \textit{macrophysical} time delays are distinct from, and in addition to, the \textit{microphysical} time delays extensively studied in the literature (see Section~\ref{sec:introduction}). The latter are usually calculated from the time of numerical BH merger and account for processes governed by unresolved sub-resolution physics.

\subsubsection{Selection cuts} \label{sec:selection-cuts}
The macrophysical time delays described above are applied to the BH merger population in \Fable{} after a few selection cuts are made. As seen in Fig.~\ref{fig:bh_masses}, the distribution of BH masses peaks at the low-mass end. These low-mass BHs are usually located in satellite subhaloes which quickly merge with a nearby, more massive subhalo, leading to a numerical BH merger. Because of our poor understanding of seeding mechanisms we only keep merger events in which both BH particles have a mass $M_{\rm BH} \geq 10^6$ \Msun{} at the time of numerical BH merger. There are a total of 91,879 numerical BH mergers in \Fable{} out of which 11,016 ($\sim 12\%$) involve two non-low mass BHs.

Our method assumes that DF on galactic scales is well resolved, ensuring that the subhalo merger time-scales in cosmological simulations are reliable~\citep{Rodriguez_Gomez_2015}. To mitigate numerical artefacts that could undermine this assumption, we focus exclusively on merger events where the constituent BHs reside in subhaloes with a total stellar mass, measured within twice their stellar half-mass radius, of at least 100 times the mass of a star particle, i.e.~$100 \times (6.4\times 10^6 \, h^{-1}$~\Msun{}), at the time of host subhalo identification. This further reduces our sample size from 11,016 down to 10,716 BH merger events.

\subsubsection{BH accretion} \label{sec:accretion}
If BH particles were to remain at the centres of their host subhaloes between their numerical merger and the eventual subhalo merger, they would continue accreting mass, leading to more massive BH at the time of merger. To account for this additional growth, we incorporate simple accretion onto the BHs in post-processing.

To do this, we replicate the simulation's accretion model as described in Section~\ref{sec:Fable_BHs}. At each snapshot between the numerical BH merger and the subhalo merger, we retrieve the physical properties of the 32 gas cells closest to the position of the subhalo's most bound particle. These properties are used to calculate the density, sound speed, and pressure in the central region where the BH particle would reside, with values interpolated for the time intervals between snapshots. The BH mass is then evolved according to the Eddington-limited Bondi-Hoyle-Lyttelton rate, as defined in equation~\eqref{eqn:accretion}\footnote{Although the $v_{\text{BH}}$ term in the Bondi-Hoyle-Lyttelton formula can suppress the BH accretion rate, we omit it, as is done in \Fable{} and many other cosmological simulations that implement BH repositioning (see Section~\ref{sec:Fable_BHs}), in order to replicate the simulation’s accretion prescription as closely as possible. We also apply the same pressure criterion as described in Section~\ref{sec:Fable_BHs}.}. Note that while this approach allows us to gain insight into the likely BH growth between numerical BH mergers and subhalo mergers, by construction this post-processing model cannot account for BH feedback which may likely affect the central subhalo properties and the rate of BH accretion itself, an issue both for the original \Fable{} simulation and our post-processing analysis. 

\subsubsection{Intervening mergers} \label{sec:intervening}

Mergers between two BH particles in \Fable{} do not happen in isolation. It is common for such merger events to occur in close proximity to other mergers within a relatively short period of time, for example, when multiple satellite subhaloes, each hosting central SMBHs, merge with a central subhalo hosting another SMBH. Removing low-mass BHs from our sample, as discussed in Section~\ref{sec:selection-cuts}, reduces the number of such events drastically, although it does not completely eliminate them. 

Incorporating macrophysical time delays can potentially alter the sequence of merger events in the simulation. This effect is illustrated in Fig.~\ref{fig:merger_process}, where a third BH merges with the remnant BH from the initial merger before the host subhaloes have fully merged. If we were to recompute the order of the mergers within our model, this third BH should merge with BH1 before the remnant of this event then merges with BH2. Such reordering could influence the total masses and mass ratios of the BHs involved in these mergers. However, only 452 out of the 10,716 merger events (4\%) that meet our selection criteria experience such a sequence reversal due to macrophysical delays. While other events may involve intervening mergers with low-mass BHs, these are not expected to notably affect the mass ratios or chirp masses of the primary merger events. Given the low frequency of intervening mergers with non-low mass BHs, we neglect this effect without significantly compromising the population-level results.

\subsection{Results: Macrophysical time delays to SMBH mergers} \label{sec:results-delays} 

In this section, we analyse the macrophysical time delays discussed in Section~\ref{sec:methods-delays} to quantify the impact of numerical effects that lead to premature BH mergers within \Fable{}.

From a total of 10,716 mergers which pass our selection criteria as described in Section~\ref{sec:selection-cuts}, only 513 merger events have no added time delay ($\sim 5\%$). In these merger events, the host subhaloes merge \textit{before} the BH particles, i.e.~the BH merger happens within the remnant host galaxy. These 513 events occur across a range of different redshifts and involve different BH masses. Although in these events, there is a trend towards more massive primary BHs of order $10^{10}$~\Msun{} residing in haloes at the more massive end of the mass spectrum, and secondary BHs residing in more massive subhaloes. This is consistent with expectations that more massive subhaloes will have a deeper gravitational potential and thus are more capable of `holding on' to their central BHs during subhalo mergers. In such cases, after the subhalo merger, the BH orbits are artificially influenced by the repositioning scheme, and microphysical delays should therefore be applied starting from the earlier time of subhalo merger rather than numerical BH merger. However, this correction would only affect a small fraction of the BH merger population and we do not model it in this work. In 541 merger events, the host galaxies merge in the snapshot directly after the numerical BH merger. 

Our time delay estimates are limited by the coarse temporal resolution of subhalo data, which is recorded only at snapshot intervals, unlike the more finely sampled BH merger data. As a result, delays are slightly overestimated, since the actual subhalo merger could occur at any time between the last snapshot where both subhaloes are resolved and the subsequent one. The snapshot spacing varies with redshift, averaging $\sim 0.1$~Gyr at $z = 3$ and $\sim 0.15$~Gyr at $z = 0$, which sets an upper bound on the timing uncertainty.

In some cases, the host galaxies of the BHs involved in a merger event fail to merge by the end of the simulation, i.e.~by $z = 0$. This happens for 3,140 events ($\sim 29\%$). For brevity, we will refer to such events as `non-mergers' from here onwards. 

\begin{figure}
    \centering
    \includegraphics[width=\columnwidth]{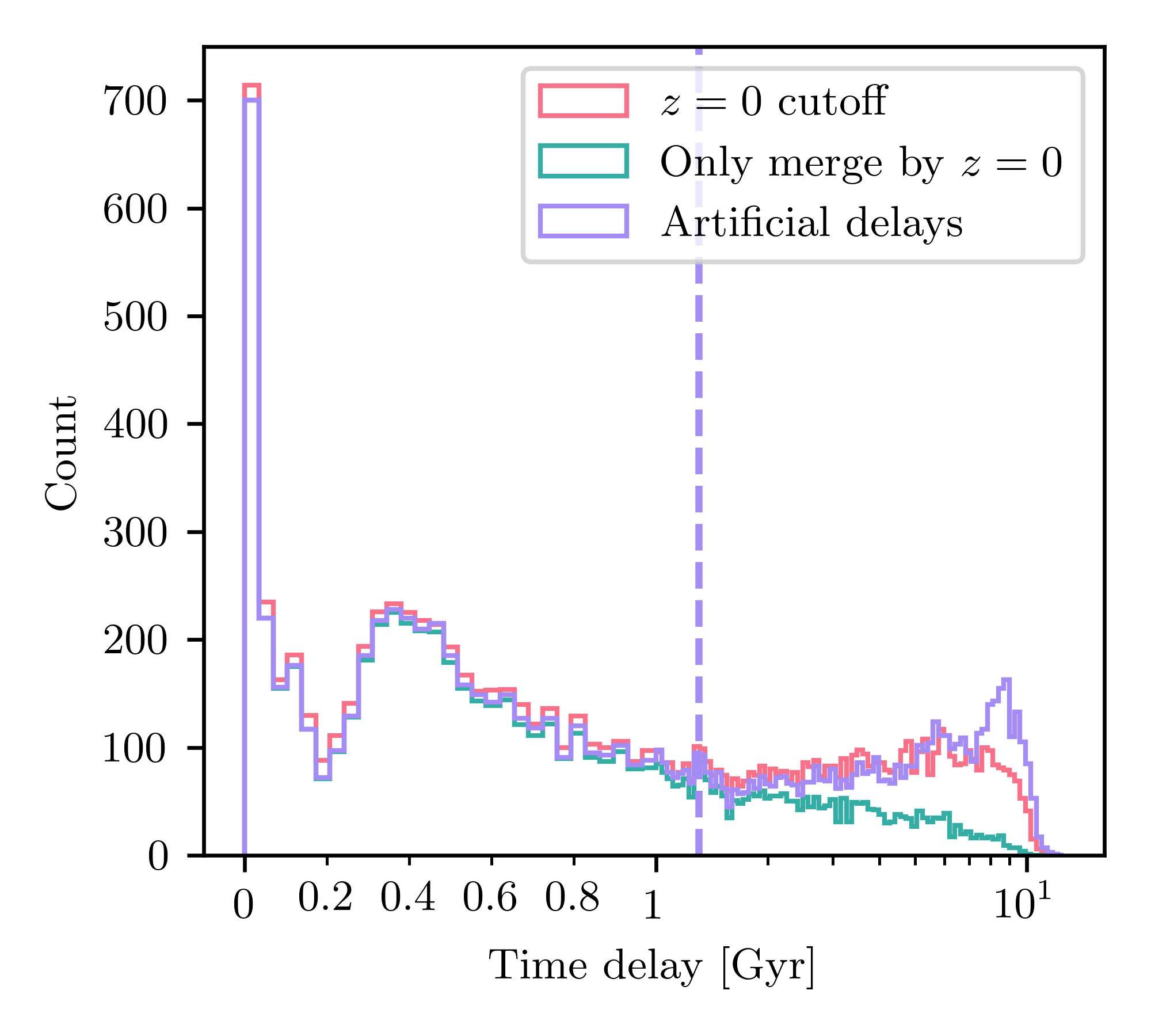}
    \caption{Distribution of macrophysical time delays in our post-processing model. The pink distribution assumes that all host galaxies merge by the end of the simulation. This is done by calculating the elapsed time until $z = 0$ for merger events whose host galaxies do not merge within the simulation. The teal distribution removes these events and only shows the time delays of BHs residing in host galaxies that merge by $z = 0$. The purple curve shows the time delays after artificially delaying those BH mergers whose host galaxies do not merge by $z=0$ as described in Section~\ref{sec:results-delays}. Time delays until 1~Gyr are shown on a linear scale, while longer time delays are shown on a logarithmic scale. The vertical dashed line shows the time delay of 1.3~Gyr for the merger event shown in Fig.~\ref{fig:merger_process} and corresponding to the median of the latter distribution.}
    \label{fig:time_delays}
\end{figure}

The distributions of macrophysical time delays in our model, with different treatments of non-mergers, are shown in Fig.~\ref{fig:time_delays}. The pink histogram shows the distribution of time delays assuming that all host galaxies merge by redshift zero, i.e.~for non-mergers, we calculate the time elapsed until redshift zero from the point of numerical BH merger. We then omit these non-merger events and show the macrophysical time delay distribution for those events whose host galaxies do merge by the end of the simulation, in teal. We use this distribution to obtain a probability distribution function and use this to estimate the total macrophysical time delay for non-mergers based on the time elapsed between the numerical BH merger and $z=0$. This distribution is then shown by the purple line. This latter distribution gives a maximum macrophysical time delay of 12.6~Gyr and a median of 1.3~Gyr. We find a weak correlation between specific gas mass fraction and our macrophysical time delays, with shorter delays associated with higher gas fractions. This likely arises because, in denser environments, the BHs have smaller smoothing lengths, reducing the likelihood of numerical BH mergers during subhalo interactions.

The merger event illustrated in Fig.~\ref{fig:merger_process} has a time delay equal to 1.3~Gyr, corresponding to the median value in our sample. This delay, indicated by the vertical purple dashed line in Fig.~\ref{fig:time_delays}, provides a representative example of the merger events in our sample. Notably, the host galaxies in Fig.~\ref{fig:merger_process} undergo significant morphological transformations between the numerical BH merger and the coalescence of the host subhaloes. The distribution of time delays in Fig.~\ref{fig:time_delays} thus suggests that these macrophysical delays drive substantial changes in the morphology of SMBH merger host galaxies. 
To compare to microphysical delays obtained from cosmological simulations, we refer to the study by \citet{Kelley_2016} in which the hardening timescales from DF, stellar scattering, gas drag and GW emission are modelled using a post-processing approach based on the properties of the remnant host galaxy. In their fiducial model, they find a median microphysical time delay for the full sample of 29~Gyr. This is much longer than the age of the Universe and only 20\% of the binaries coalesce before $z=0$. The median is reduced to 6.9~Gyr when considering only systems with a total mass $> 10^8 $~\Msun{} and mass ratio $ > 0.2$.
Although these values strongly depend on model assumptions such as the loss cone refilling rate, they are comparable to our macrophysical delays, highlighting the importance of considering both effects in predictions for future observations.

The suppression between the first two distributions (pink and teal) in Fig.~\ref{fig:time_delays} is directly caused by non-merger events, by definition of the distributions. While a slight suppression is present across the entire range of time delays, it becomes most pronounced for delays exceeding $\sim 1$~Gyr. The consistent suppression across the full time range corresponds to premature BH mergers during ongoing galaxy mergers, as highlighted in Fig.~\ref{fig:merger_process}, that do not merge by $z=0$, resulting in non-merger events. This effect is particularly relevant at lower redshifts. However, the majority of these non-mergers, which dominate the tail-ends of the distributions, stem from the numerical effect illustrated in Fig.~\ref{fig:merger_process_2} in which BH particles are repositioned to different subhaloes and lose their association with their original host subhaloes. Consequently, these hosts often evolve independently, and are likely to never merge, leading to large macrophysical delays in our model.

\begin{figure}
    \centering
    \includegraphics[width=\columnwidth]{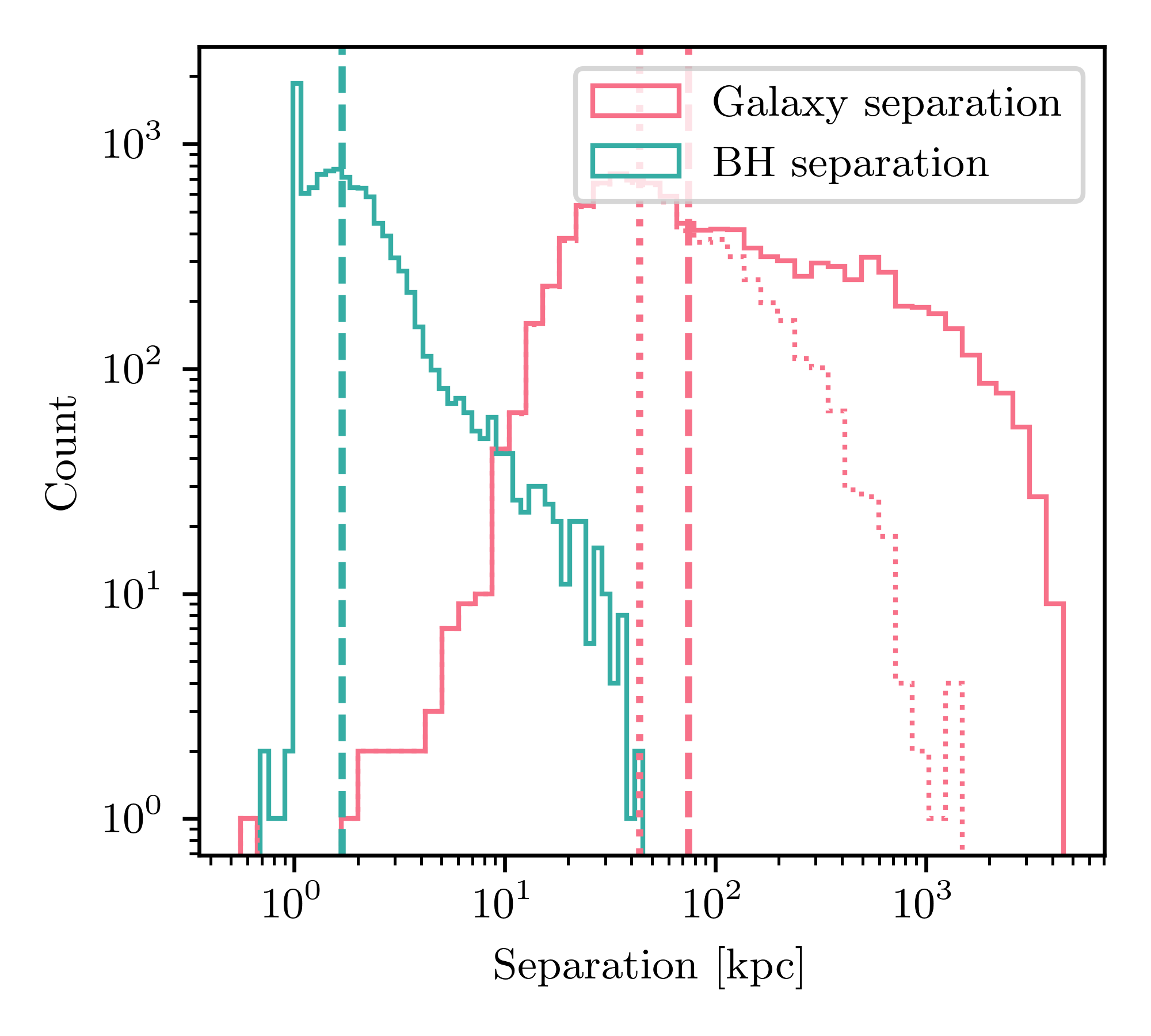}
    \caption{Separations of BHs calculated using the smoothing length of the more massive BH in teal, compared to the host galaxy separations in the snapshot preceding the host-galaxy merger in pink. The solid pink distribution includes the separation of the host subhaloes at $z=0$ for non-mergers. The dotted pink line plots the distribution of host galaxy separations without these non-merger events. The vertical lines show the median of the distributions, with the dashed lines corresponding to the two former distributions (in matching colours) and the dotted line corresponding to the events without non-mergers.}
    \label{fig:separations}
\end{figure}

To calculate time delays from sub-grid hardening mechanisms, as described in Section~\ref{sec:introduction}, an initial separation at which to initialise these sub-resolution prescriptions is required. This should roughly match the separation of the BHs at which their orbits within the cosmological simulation are no longer tracked, i.e.~the separation at which they are considered to be merged within the simulation. In previous works, an upper limit for this initial separation was taken to be the smoothing length of the more massive BH particle in the snapshot directly preceding the numerical merger event~\citep{Kelley_2016}, or a small multiple of the spatial resolution of the simulation~\citep{kunyang_2024}. We plot the distribution of BH separations using the particle smoothing lengths for our sample of merger events in \Fable{} in teal in Fig.~\ref{fig:separations}, which matches distributions shown by \citet{Kelley_2016}. However, since the time of BH merger in \Fable{} is dominated by numerical effects, as depicted in Figs.~\ref{fig:merger_process} and \ref{fig:merger_process_2}, the corresponding BH separations at merger are also plagued by these numerical artefacts, and no longer correspond to the particle smoothing lengths at the redshift at which we consider the BHs to be finally merged within our macrophysical delay model. 

We argue that a better, albeit more conservative estimate for the initial separation which can be obtained from these large-volume cosmological boxes, is the separation between the original host subhaloes, in the snapshot before subhalo merger. The distribution of these separations for the same sample of merger events is shown in pink in Fig.~\ref{fig:separations}, with the solid line corresponding to the full merger sample and the dotted line excluding non-merger events. The corresponding vertical dashed lines in Fig.~\ref{fig:separations} indicate the median for each distribution, with the median galaxy separation being roughly two orders of magnitude larger than the BH separation set by their smoothing lengths. This would imply that a longer phase of orbit shrinking via DF is required when modelling sub-resolution hardening mechanisms using the galaxy separation as an educated guess for the initial separation. This would inadvertently lead to {\it significantly} longer microphysical time delays than our current best estimates when using BH smoothing lengths, compounding the added macrophysical delays from our model.

\subsubsection{Implications for GW observations} \label{sec:res_implications}

In order to quantify the effects of our macrophysical delays on GW observations, we compare the subsample of merger events passing our selection criteria in \Fable{} without any modifications, with added macrophysical time delays but no modification to their masses, and with added delays while allowing for growth by accretion to happen as discussed in Section~\ref{sec:accretion}. In order to isolate the effects of our macrophysical delays on the observable population of SMBH mergers, we neglect any microphysical time delays arising from unresolved hardening mechanisms, unless otherwise stated.

\begin{figure*}
    \centering
    \includegraphics[width = 2\columnwidth]{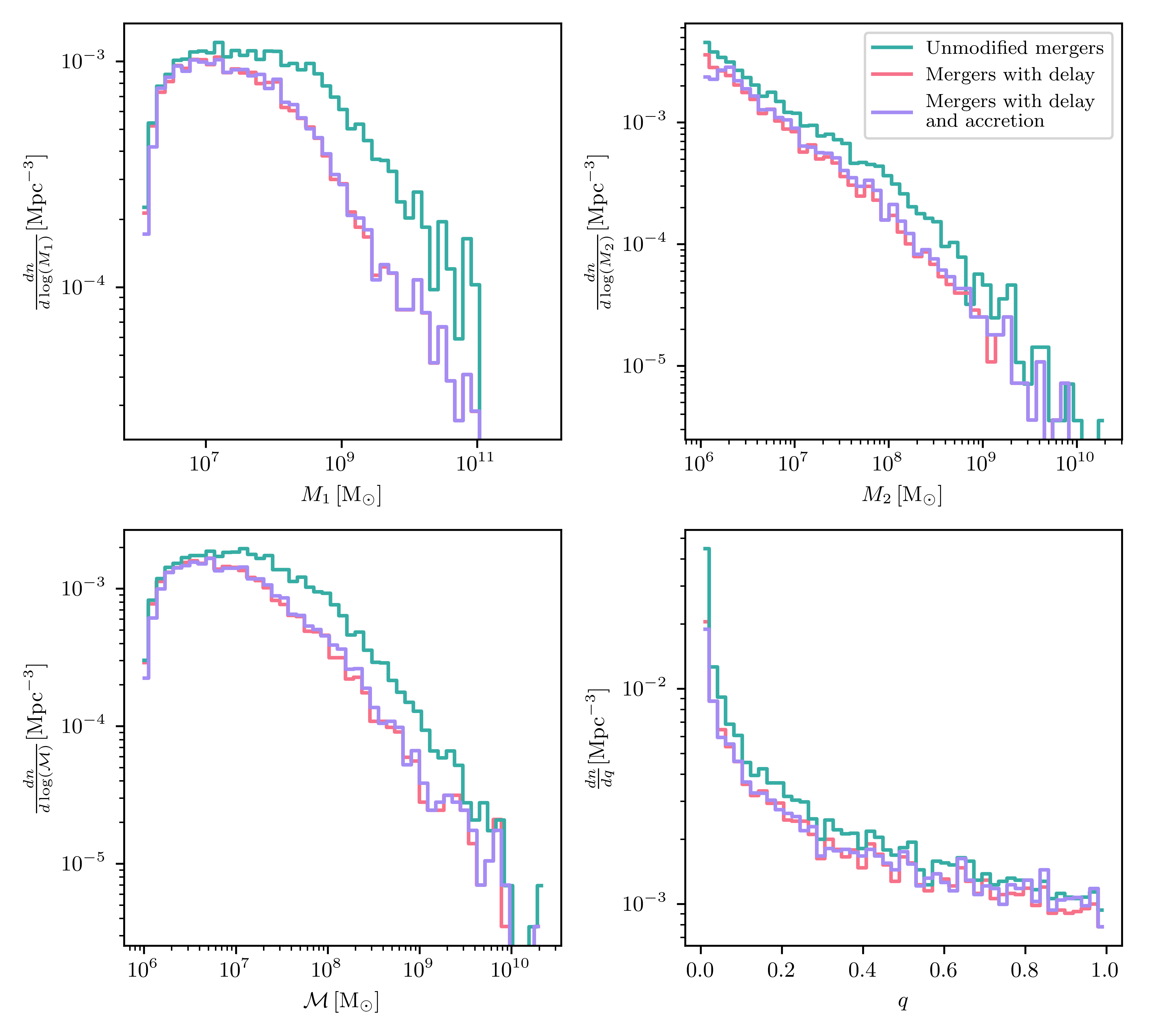}
    \caption{Distributions of the number density of events as a function of primary mass $M_1$ (top left), secondary mass $M_2$ (top right), chirp mass $\mathcal{M}$ (lower left) and mass ratio $q$ (lower right). The teal lines correspond to the unmodified merger population in \Fable{}, the pink lines add our macrophysical delays without changing the BH masses, and the purple lines show the distribution when accounting for BH accretion during the macrophysical delays.}
    \label{fig:mass_histograms}
\end{figure*}

As indicated in Fig.~\ref{fig:time_delays}, in our model the number of merger events is reduced, as some host galaxies do not merge by the end of the simulation. Consequently, introducing macrophysical delays alters the mass distribution of the merger population, with accretion further modifying their masses. Fig.~\ref{fig:mass_histograms} displays the number density of merger events occurring before $z=0$ as a function of key mass parameters: primary mass $M_1$, secondary mass $M_2$, chirp mass $\mathcal{M}$, and mass ratio $q = M_2/M_1$, where the chirp mass is defined as
\begin{equation}
    \mathcal{M} = \frac{\left(M_1 M_2\right)^{3/5}}{(M_1+M_2)^{1/5}}\,.
\end{equation}
The distribution with respect to primary mass (top-left) shows that while the number of mergers decreases across all masses after introducing macrophysical delays, the suppression is strongest at high $M_1$, which is then reflected in the chirp mass distribution (bottom-left). This effect arises because SMBH growth through accretion and mergers leads to the formation of larger BHs at later cosmic times in \Fable{}, as highlighted by the BHMF in Fig.~\ref{fig:bh_masses}. Since high-mass mergers predominantly occur at lower redshifts, applying time delays shifts many of these mergers into the future, removing them from the observed population. In contrast, the suppression in secondary mass $M_2$ (top-right) is more uniform across the full mass range. The reduction in high-mass primaries also decreases the number of mergers with the most extreme mass ratios, though the overall number of merger events declines across all $q$ values (bottom-right). Notably, incorporating accretion during the added macrophysical delays does not significantly alter any of the distributions in Fig.~\ref{fig:mass_histograms}.

\begin{figure}
    \centering
    \includegraphics[width=\columnwidth]{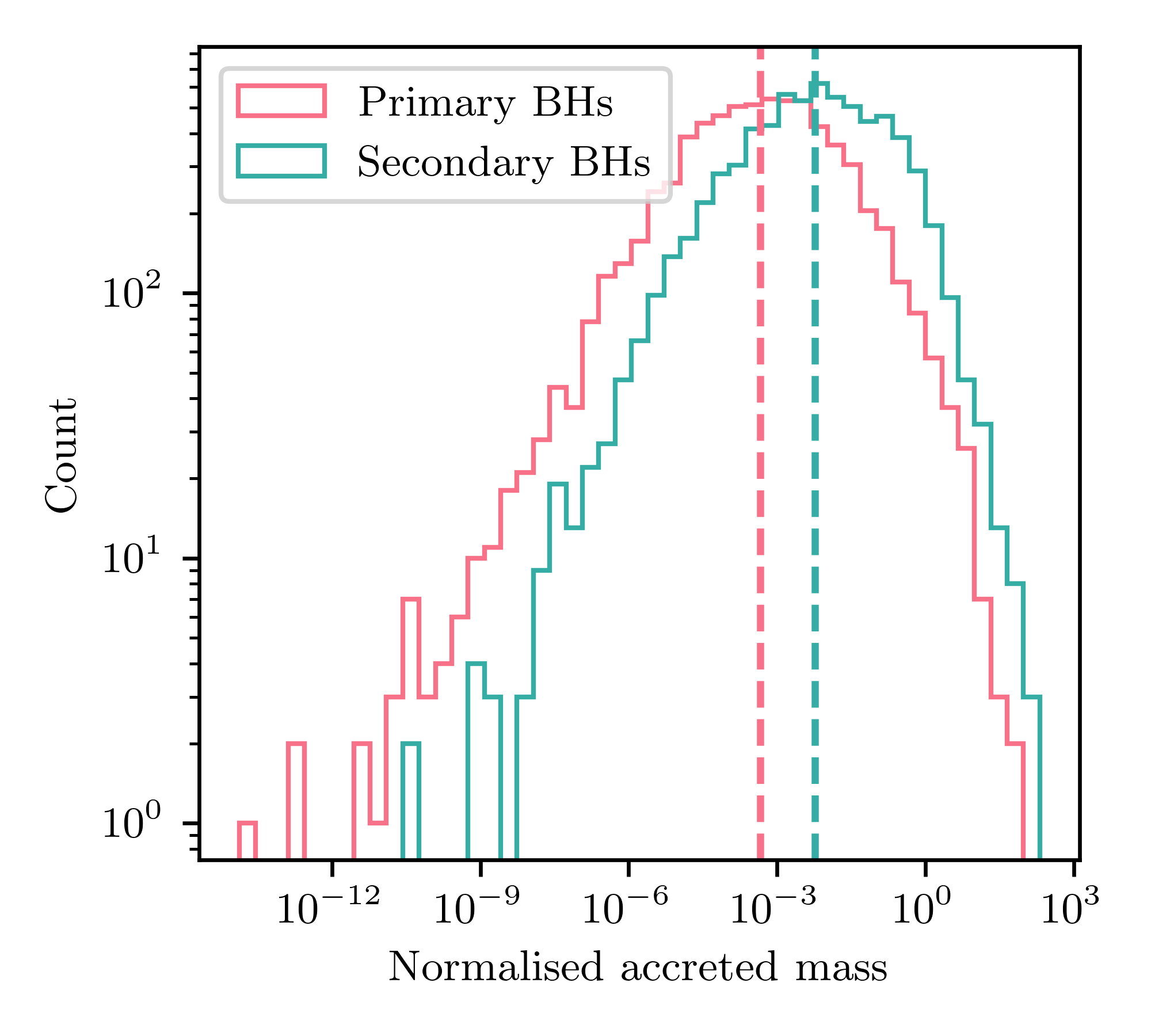}
    \caption{Normalised accreted mass between the numerical BH merger and the end of the macrophysical time delay, i.e.~$(M_{\rm BH,merger+\Delta t_{\rm macro}} - M_{\rm BH,merger})/M_{\rm BH,merger}$ (see main text for more details), for the subset of BH merger events whose host galaxies merge by $z=0$. The distribution for primary BHs is shown in pink, while that for the secondary BHs is shown in teal. The corresponding vertical dashed lines indicate the median of each distribution. This shows that secondary BHs systematically accrete more mass than primary BHs in our post-processing approach.}
    \label{fig:normalised_mass}
\end{figure}

We further investigate this accreted mass during our macrophysical delays in Fig.~\ref{fig:normalised_mass}. This figure shows the distribution of normalised accreted mass for merger events whose host galaxies merge by $z=0$. Normalised accreted mass is defined as $(M_{\rm BH,merger+\Delta t_{\rm macro}} - M_{\rm BH,merger})/M_{\rm BH,merger}$, where $M_{\rm BH,merger}$ is the BH mass at the time of numerical BH merger, while $M_{\rm BH,merger+\Delta t_{\rm macro}}$ is the BH mass after the macrophysical time delay. We exclude BHs which accrete no mass during the macrophysical time delay, either due to this delay being equal to zero, or because they reside in gas-poor galaxies. We find that the amount of mass accreted during this delay is generally low. However, about 4\% of the BHs in this figure at least double their mass, likely corresponding to the tail of the macrophysical time delay distribution (teal colour), showing BHs that merge by $z=0$, in Fig.~\ref{fig:time_delays}. We also find that secondary BHs systematically accrete more mass than primary BHs during these delays. As discussed in Section~\ref{sec:accretion}, this happens because our post-processing approach cannot self-consistently model BH feedback. Secondary host galaxies which are prematurely stripped of their central BH particle (see Fig.~\ref{fig:merger_process}) will not have any BH feedback to regulate central gas densities, allowing BHs to accrete at higher rates in post-processing. On the other hand, primary host galaxies which tend to host the more massive remnant BH during this delay than should be the case, experience feedback that is too strong (due to BH mass dependence), pushing gas away from the central regions of the galaxy, leading to lower accretion rates calculated in post-processing. This might also affect the SFRs in these host galaxies. 

\begin{figure}
    \centering
    \includegraphics[width=\columnwidth]{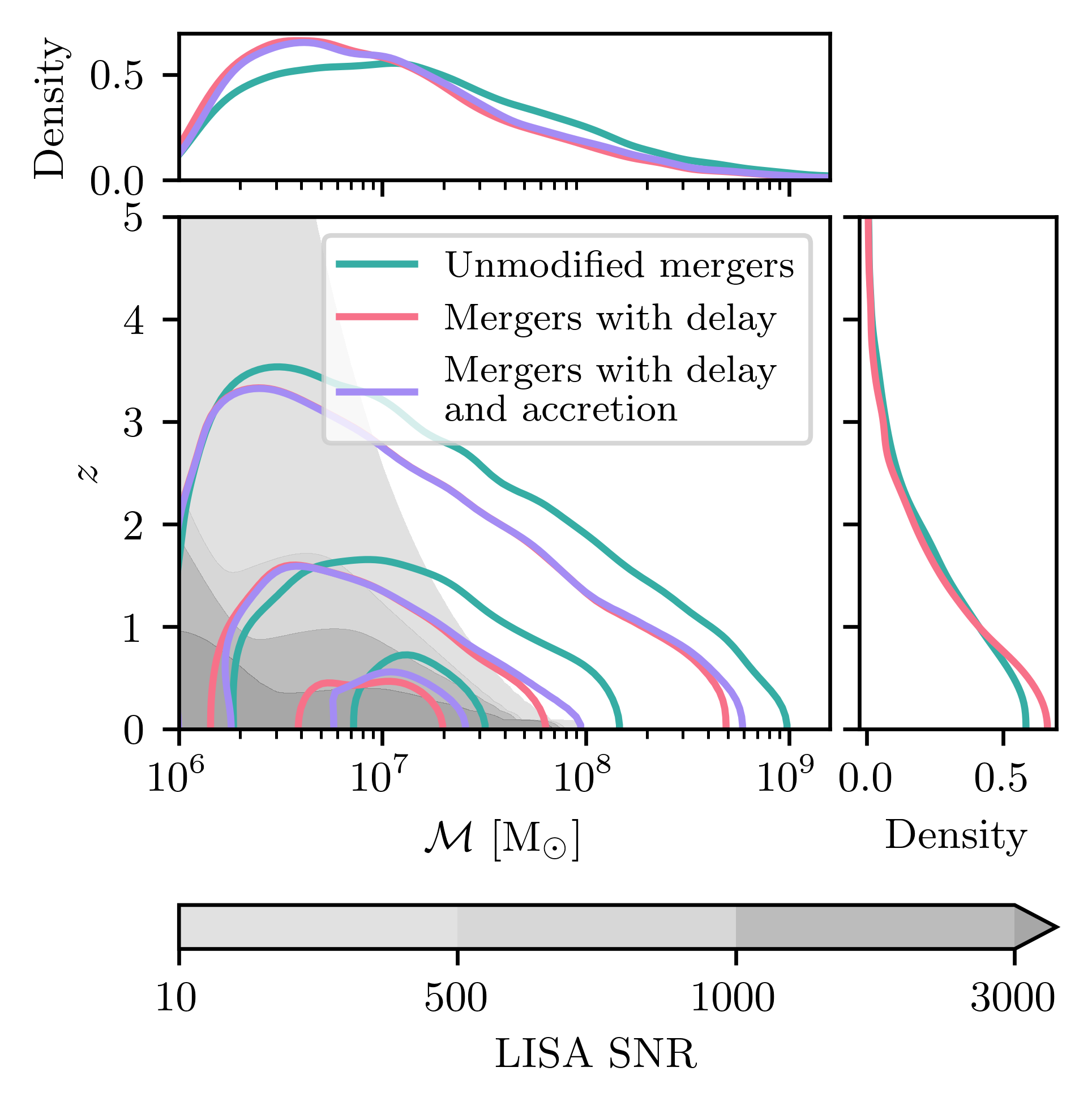}
    \caption{The main figure shows the distribution of merger events in redshift-chirp mass space. The teal contours show this distribution from \Fable{}, the distribution shown in pink adds our macrophysical time delays in post-processing whilst leaving the BH masses unchanged, and the distribution shown in purple adds the same macrophysical time delays whilst also allowing the BHs to accrete during this delay time. The contours show 10\%, 50\% and 90\% of our sample. Grey shaded regions show the curves of constant SNR for \textit{LISA} as indicated by the colourbar and as discussed in Section~\ref{sec:res_implications}. The top plot shows the marginal distribution of chirp masses while the right hand side plot shows the distribution of redshifts.}
    \label{fig:contour}
\end{figure}

Fig.~\ref{fig:contour} illustrates the relation between merger chirp mass $\mathcal{M}$ and redshift $z$ in \Fable{} for our three different models. Note that the teal contours, corresponding to the unmodified population in \Fable{}, contain a larger number of events than the contours including delays, as the latter do not include non-merger events, making them a subset of the original population. In the top and right sub-panels the marginal distributions of chirp mass and redshift, respecitvely, are shown.

Our macrophysical delays slightly suppress events at redshifts $z >1$, with the effect of increasing the number of events occurring at $ 0 < z < 1$, although the location of the peak of the redshift distribution does not shift significantly. A more pronounced effect is the shift toward lower chirp masses across all redshifts which can be seen both in the 2D and 1D distributions. The broad peak in the distribution of chirp masses at $ \sim 10^7$ \Msun{} produced by \Fable{} is shifted towards $ \sim 3 \times 10^6$ \Msun{} and becomes more pronounced. The teal contours from the unmodified \Fable{} simulation in the 2D contour plot highlight how high-mass mergers occur at later redshifts. Showing the joint distribution of chirp mass and redshift breaks the degeneracy seen in the top panel and in Fig.~\ref{fig:mass_histograms} which group mergers across all redshifts. Because of this, it is clearer to see the impact of accretion on the distribution of chirp masses at specific redshifts. In general, our accretion model shifts mergers at fixed $z$ values towards higher chirp masses, however this effect is not large enough to recover the highest chirp mass events that are lost with the introduction of macrophysical time delays. 

In the shaded, gray background of Fig.~\ref{fig:contour} we plot the contours of constant \textit{LISA} signal to noise ratio (SNR), i.e.~the ``waterfall plot''. The SNR calculations were performed using BOWIE \citep{Katz_2018} and the IMRPhenomD waveform~\citep{Husa_2016, Khan_2016b}. For these SNR calculations we assume two non-spinning BHs with a fixed mass ratio of $q=0.5$ on a circular orbit and use the sky-averaged \textit{LISA} Phase A noise curve above $10^{-4}\,\mathrm{Hz}$, including the confusion noise from galactic white dwarf binaries~\citep{Ruiter_2010, lisa_redbook}. Despite these simplifications, these sensitivity curves give a good impression of \textit{LISA}'s sensitivity and show the implications of our macrophyiscal time delays. For the high-mass binaries studied here, \textit{LISA} will observe the entire GW signal above $10^{-4}\,\mathrm{Hz}$ which lasts $\lesssim 1$ month. \textit{LISA} will also observe SMBH mergers with smaller chip masses which are not included here because they do not pass the mass cut applied to the \Fable{} catalogue described in Section~\ref{sec:selection-cuts}. The shift in the SMBH population towards lower redshifts due to macrophysical delays means that the binaries that merge will have higher \textit{LISA} SNRs than predicted by \Fable{}, improving their detectability. Furthermore, the suppression of high-mass mergers implies that \textit{LISA} will be able to detect a larger percentage of merger events at higher SNR, leading to a more complete understanding of the merging SMBH population. 

\begin{figure}
    \centering
    \includegraphics[width=\columnwidth]{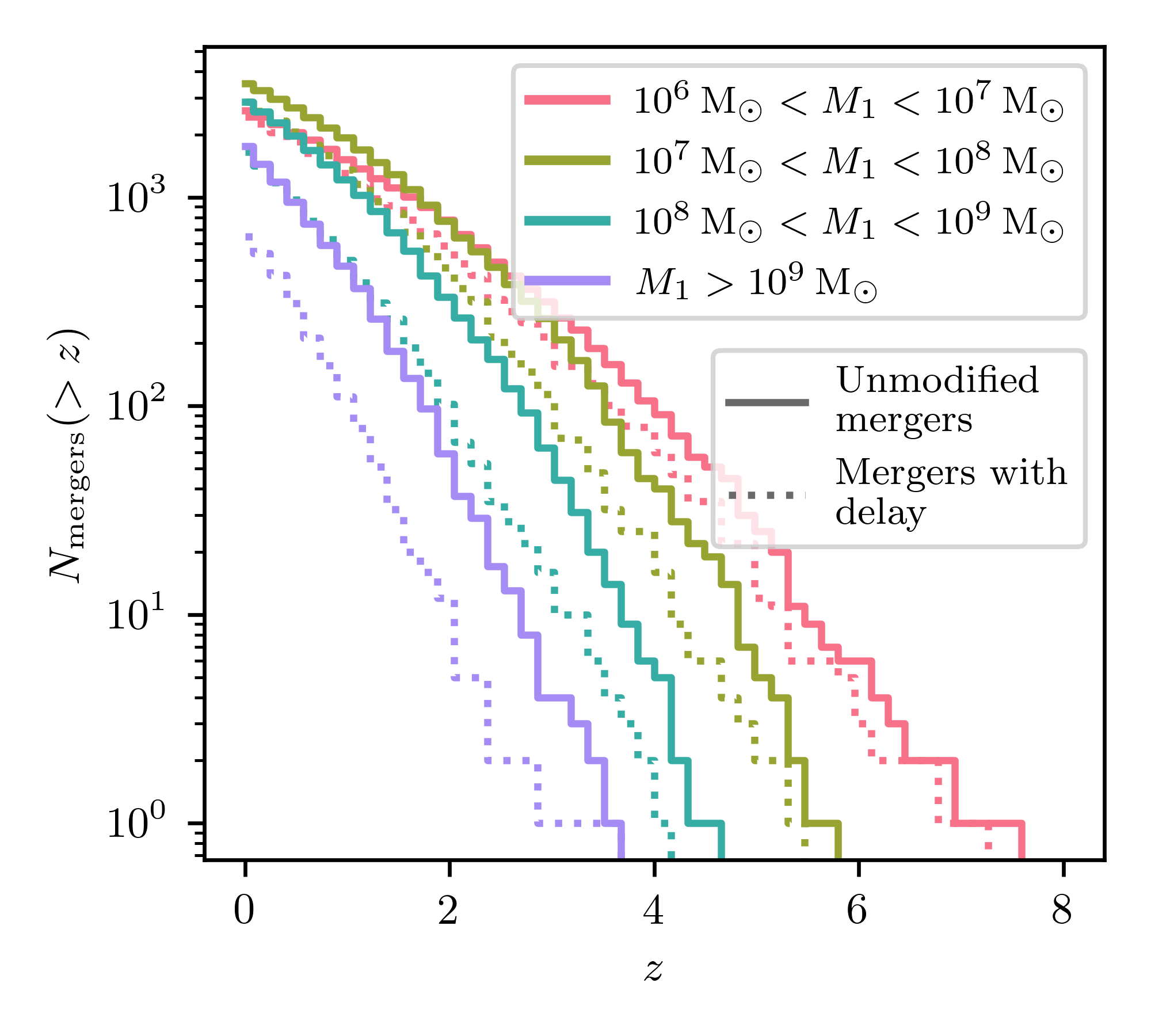}
    \caption{Cumulative merger number binned by primary mass. The solid lines show the unmodified merger population in \Fable{}, whilst the dotted lines show the cumulative distribution after adding macrophysical time delays.}
    \label{fig:cumulative}
\end{figure}

The suppression of high-mass merger events is also highlighted in Fig.~\ref{fig:cumulative} which shows the cumulative number of mergers as a function of redshift binned according to the primary mass, $M_1$. The unmodified merger population in \Fable{} is compared to the merger population after the addition of our macrophysical time delays. Note that accounting for accretion does not significantly alter the BH masses, and therefore does not affect the binning or the results shown in this figure. The number of mergers is reduced across all mass bins with the introduction of these time delays although the effect becomes more noticeable as the mass of the primary increases, in agreement with Figs.~\ref{fig:mass_histograms} and \ref{fig:contour}. In both cases, BH mergers with $10^7 \, \Msun{} < M_1 < 10^8$ \Msun{} dominate the merger rate which is broadly consistent with previous predictions from cosmological simulations~\citep{Salcido_2016, DeGraf_2019}. At higher redshifts, the number of mergers increases with decreasing primary mass which is consistent with the cosmic growth of SMBHs via accretion and mergers. 

\begin{figure}
    \centering
    \includegraphics[width=\columnwidth]{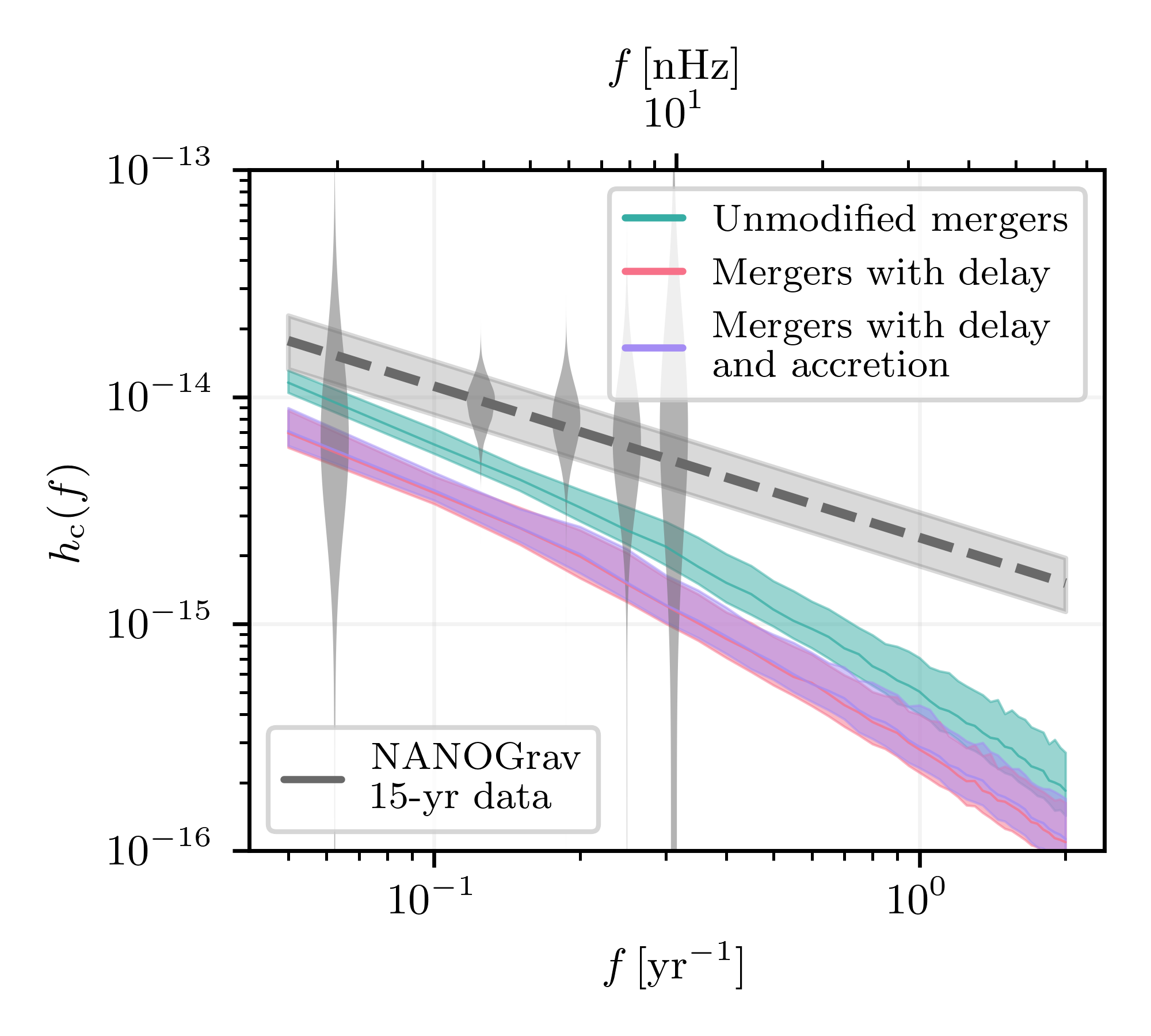}
    \caption{The SGWB strain generated from the subsample of merger events in \Fable{} passing our selection criteria (see Section~\ref{sec:selection-cuts}). The unmodified population is shown in teal, the population with added macrophysical time delays is shown in pink, while the population with added macrophysical time delays as well as allowing for BH accretion during these delays is shown in purple. For all simulation models we further assume that unresolved hardening microphysical delays are equal to 1~Gyr. For the unmodified merger population we use the initial separations set by the BH smoothing lengths whereas for the two populations with added macrophysical delays we initialise the binaries at the separation of the host galaxies. We draw 1000 realisations from our discrete binary population in order to estimate the SGWB in each case. The solid lines indicate the median of these realisations, while the interquartile range is highlighted by the shaded regions. The grey dashed line with the shaded region shows the power law fit from the NANOGrav 15-year data set (median with 90\% credible interval). The grey violins show the Hellings-Downs correlated free spectrum posteriors from the same data set for the first five frequency bins~\citep[see][for more details]{agazie_2023}. The frequency range plotted corresponds to the sensitivity of PTAs after twenty years of observations.}
    \label{fig:gwb}
\end{figure}

Fig.~\ref{fig:gwb} shows the expected SGWB sourced by the population of merger events in \Fable{} passing our selection criteria within the PTA frequency sensitivity range for all three of our models. To calculate the strain we use the \texttt{holodeck} package for SMBH binary population synthesis and GW calculations~\citep{Kelley_2016, agazie_2023}. Based on each population of mergers, discrete realisations of the binary population are generated using a weighted sampling approach, with each SMBH binary in our sample being weighted using a Poisson distribution. For each realisation, the total strain in each observed frequency bin is calculated by averaging over the angle and polarisation for binaries on circular orbits~\citep{finn_2000}.
Accounting for unresolved hardening processes is expected to significantly further delay SMBH mergers, but their modelling still involves large uncertainties. To isolate the effect of our macrophysical delays, we adopt a fixed microphysical hardening time delay of 1~Gyr as a simplifying assumption, following the approach by~\citet{agazie_2023}. We set the initial separations of the unmodified merger events to the particle smoothing lengths of the primary BHs (teal distribution in Fig.~\ref{fig:separations}). For the two populations with added macrophysical delays we use the host galaxy separations (pink distribution in Fig.~\ref{fig:separations}) as the initial separations.

As expected, the addition of macrophysical delays suppresses the SGWB across all frequencies due to the reduction in the number of merger events that occur before $z=0$, specifically at the high-mass end of the distribution, as illustrated in Fig.~\ref{fig:cumulative}, which is the main contribution to the SGWB. Allowing the BHs to grow via accretion during the added macrophysical delays does not significantly boost the amplitude of the signal. Theoretically, for a large population of sources, and assuming circular orbits decaying primarily due to GW emission, the characteristic strain is expected to scale with frequency as $h_c(f) = A_\mathrm{GWB} \left(f/f_0\right)^{-2/3}$~\citep{phinney_2001}. We show this as the grey dashed line in Fig.~\ref{fig:gwb} using the fiducial result from the NANOGrav power-law analysis of the 15-year dataset with an amplitude at the reference frequency of $A_\mathrm{GWB} = 2.4^{+0.7}_{-0.6}\times 10^{15}$. We also show the results  of the Hellings-Downs correlated free-spectrum analysis, independent of the power-law assumption, for the first five frequency bins of the NANOGrav 15-year dataset, which give the best constraints on the SGWB amplitude~\citep{agazie_2023}.  All three of our models under-predict the amplitude of the SGWB (with respect to both the power-law and free-spectrum analyses) in the frequency range $(\sim 2-10)\,\mathrm{nHz}$ where NANOGrav is most sensitive.

This is consistent with previous predictions from cosmological simulations (see for example \citet{Kelley_2016}) and studies that suggest that BH mass relations need to have a significantly higher normalisation, or significantly evolve with redshift, in order to match the PTA signal~\citep{chen_2023b}. Reducing the fixed hardening timescale to 0.1~Gyr, as also studied by \citet{agazie_2023}, does not reconcile the predicted signal with the NANOGrav result. This highlights the necessity of using a more realistic hardening model for microphysical delays, opposed to the fixed hardening time used here, which can significantly impact the predicted amplitude of the SGWB. 

We also note the possible impact that a limited box size may have on these predictions. As BHs at the massive end of the mass spectrum with $M_\mathrm{BH} \gtrsim 10^{10}$~\Msun{}, which can have a significant impact on the amplitude of the SGWB, are missing from the cosmological box with side-length of $100 \, h^{-1}$~cMpc used for this study, due to their low number density. We have investigated the impact of adding such high-mass
BHs to the \Fable{} population and found that their inclusion can amplify the predicted SGWB signal at the low end of the frequency range, but that this amplification depends on the details of the modelling choices made for this high-mass population. A detailed investigation of the effect of the limited box size on predictions of the SGWB is left for future work.

Eccentric orbits, expected at large binary separations but not modelled here, could boost the SGWB amplitude by shortening binary lifetimes and increasing merger rates~\citep{Kelley_2017, Mannerkoski_2022}. On the other hand, if binaries remain eccentric up until the GW-dominated stage of the binary evolution, this leads to a suppression of the GW signal due to faster orbital decay, which would further increase the discrepancy with observations~\citep{Sesana_2013, chen_2017}.

Focusing on the shape of the spectrum, all three of our models deviate from the expected $f^{-2/3}$ power-law at the high-frequency end. This is expected, as at the high-frequency end the population is incomplete, with higher-mass binaries merging before reaching this frequency~\citep{sesana_2008, sato_2024}. At the low-frequency end, all three models slightly diverge from this simple power-law due to environmental effects since the inspiral at large separations is not purely driven by GW radiation. However, this behaviour is highly sensitive to the chosen sub-grid hardening model, and can also be affected by eccentric orbits~\citep{chen_2017}.

\subsubsection{Implications for EM observations} \label{sec:EM_implications}

The addition of macrophysical time delays not only impacts GW observations, as discussed earlier, but also influences key astrophysical properties of the SMBH population relevant for EM observations. 

\begin{figure}
    \centering
    \includegraphics[width=\columnwidth]{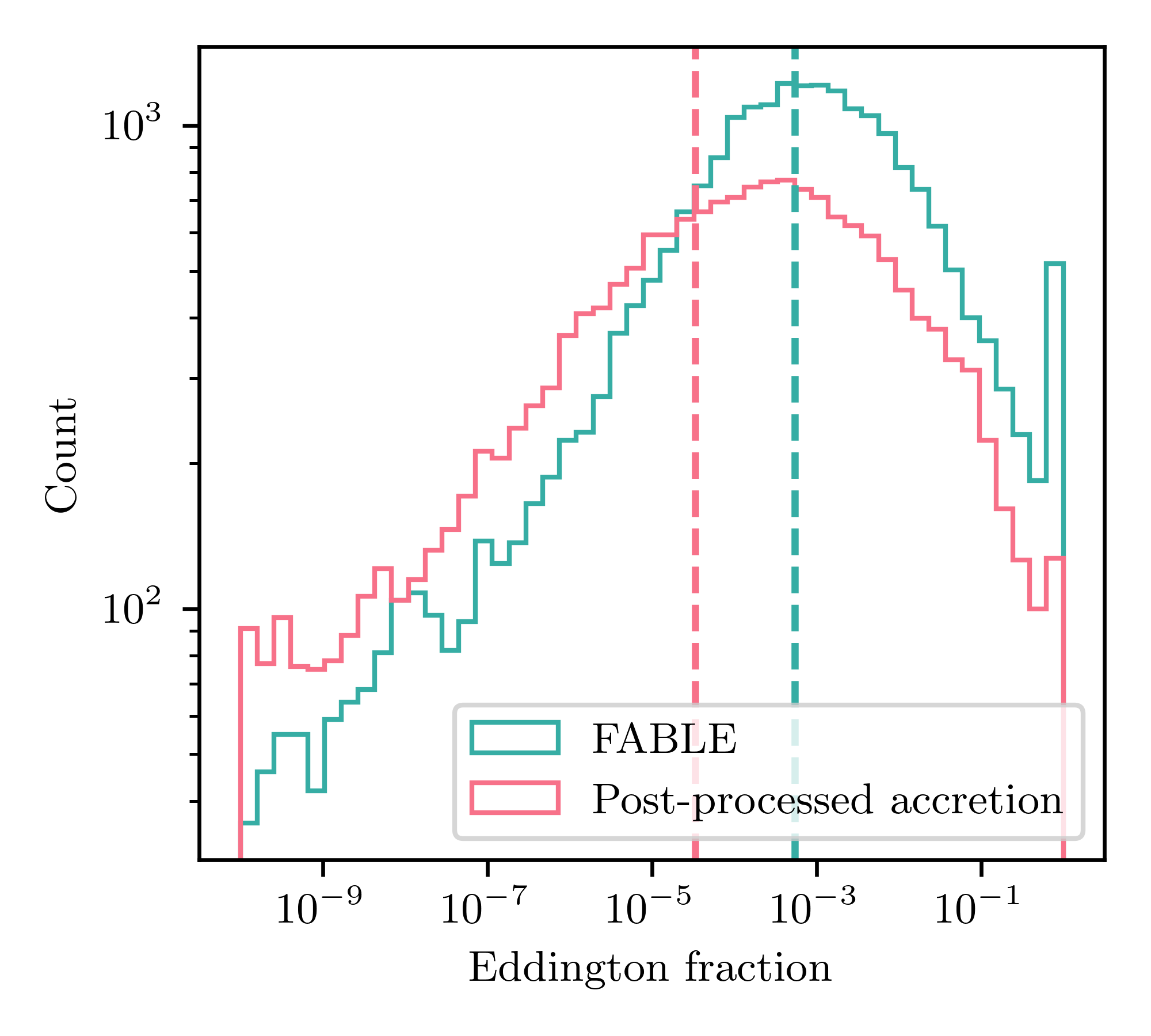}
    \caption{Distribution of Eddington fractions for dual AGN calculated in the snapshot before numerical BH merger for both the primary and secondary BHs in \Fable{} (teal line) compared to our model with macrophysical delays (pink line) where Eddington fractions are evaluated in the snapshot before host subhaloes merge. The vertical dashed lines show the median of the corresponding distribution.}
    \label{fig:edd_fractions}
\end{figure}

To illustrate this point, the distributions of Eddington fractions, as defined in Section~\ref{sec:Fable_BHs}, are shown in Fig.~\ref{fig:edd_fractions}. These distributions show Eddington fractions of dual AGN, as we group together the Eddington fractions of both the primary and secondary BHs involved in a merger event. To obtain the distribution from the unmodified \Fable{} simulation, we use the instantaneous accretion rate of the BHs in the snapshot directly preceding the numerical BH merger. In our post-processed accretion model, we use the accretion rate calculated from gas cells in the central regions of the host subhaloes as described in Section~\ref{sec:accretion}, in the snapshot before the subhalo merger. For merger events whose host galaxies do not merge by $z = 0$, we take the accretion rate in the last snapshot of the simulation. Note that the Eddington ratios of these dual AGN are measured when the BHs are at separations as shown in Fig.~\ref{fig:separations}.

In our model, the median Eddington fraction of dual AGN decreases by around one order of magnitude compared to \Fable{}. This aligns with the observed redshift evolution of Eddington fractions. At high redshifts, the central regions of galaxies contain abundant cold gas that efficiently accretes onto BHs, whereas at lower redshifts, star formation and feedback processes suppress gas inflow, reducing accretion rates and lowering Eddington fractions~\citep{shen_2012}. By delaying BH mergers, macrophysical time delays shift these events to lower redshifts, naturally leading to lower Eddington fractions. The difference in medians between the two distributions in Fig.~\ref{fig:edd_fractions} is comparable to the change in median Eddington fraction between $z=1$ and $z=2$ in \Fable{}, which corresponds approximately to the median macrophysical delay of 1.3~Gyr. Additionally, since our model incorporates macrophysical time delays on galactic scales, resulting in a higher number of binary SMBHs that do not merge by $z=0$, we expect an increased prevalence of dual AGN at lower redshifts. In summary, our model predicts a population of dual AGN with lower luminosities, at lower redshifts and with larger separations than that inferred directly from \Fable{}.

Note that the peak at high Eddington fractions in the \Fable{} distribution (Fig.~\ref{fig:edd_fractions}) arises because we are specifically considering BHs hosted in galaxies undergoing a merger event or close encounter. During galaxy mergers, strong tidal forces remove angular momentum from gas, driving it towards the central SMBHs and triggering AGN activity, which enhances accretion rates and increases Eddington fractions~\citep{Toomre_1972, Barnes_1991, Barnes_1996, talbot_2024}. In contrast, this peak is significantly less pronounced in our post-processed model, however it is not straightforward to isolate a single physical interpretation of this result. As highlighted in Fig.~\ref{fig:normalised_mass} and the corresponding discussion, more gas infalls onto the secondary BH, since no BH particle is present in the original simulation and thus no AGN feedback is available to regulate gas inflow into the centre of the subhalo. However, the presence of an over-massive remnant BH in the primary halo may have the opposite effect on the accretion rate of the primary BH in our model, suppressing it through excessive feedback\footnote{Even though these primary BHs may grow more since the accretion rate scales as $M_\text{BH}^2$.}. This limitation underscores the challenges of implementing macrophysical time delays and accretion corrections in post-processing to account for premature BH mergers. In such models, AGN feedback cannot be consistently tracked, which in turn alters the accretion history of the constituent BHs together with key host galaxy properties. 

\section{Discussion \& Conclusions} \label{sec:conclusion}

In this work, we use the \Fable{} cosmological boxes with periodic lengths of $40 \, h^{-1}$~cMpc and $100 \, h^{-1}$~cMpc~\citep{Henden_2018, Henden_2019, Henden_2019b, bigwood_2025} to study the population of merging SMBHs. We begin by examining the key properties of the BH population within the simulation and validating them against observations. Our analysis shows that \Fable{} reproduces the observed present-day BHMD and AGN luminosity function very well. However, it underpredicts the number density of low mass BHs and overpredicts the number density of the most massive BHs compared to the observationally inferred BHMF. Additionally, \Fable{} reproduces the $M_{\text{BH}} - M_\text{bulge}$ relation well, but tends to underestimate BH masses for low mass galaxies (i.e.~$M_{\rm star,HM} \lesssim 5 \times 10^{10}$~\Msun{}). These results put \Fable{} on par with other widely used cosmological simulations, with no simulation currently being able to accurately predict all observed SMBH population properties~\citep{Habouzit_2021}. 

Focusing on the population of merging BHs, we have found that the majority of BHs merge prematurely in relation to the merger of their host galaxies in \Fable{} due to two compounding numerical effects.
\begin{enumerate}
    \item In the simulation, two BHs merge when they come within a smoothing length of each other, irrespective of whether they are gravitationally bound or not. This can occur while their host subhaloes are still interacting (e.g. at the first passage) but have not yet fully merged.
    \item During subhalo interactions, strong variations in gravitational potentials, or steep potentials of nearby subhaloes can cause BHs to be gradually `repositioned’ to a subhalo other than their original host.
\end{enumerate}
These numerical effects are also expected to play a role in other widely-used cosmological simulations that implement a similar repositioning scheme as is used in \Fable{}. These include Illustris, IllustrisTNG, MilleniumTNG, EAGLE, Flamingo and Bahamas, some of which have been used in previous works to make predictions of SMBH merger rates~\citep{Kelley_2016, mccarthy_2016, Salcido_2016, DeGraf_2019, Siwek_2020,Li_2022, Pakmor_2023, Schaye_2023}. We have conducted the first study that quantifies these numerical effects and provide an initial attempt at correcting for them. We do so by adding macrophysical time delays that correspond to the time elapsed in the simulation between numerical BH merger and the merger of the host subhaloes, which we track throughout the merger process. During these delays, BHs are allowed to grow in post-processing via Eddington limited, Bondi-like accretion based on the central gas properties of their host subhaloes. Our main findings are summarised below.
\begin{itemize}
    \item These macrophysical time delays affect 93\% of the BH merging population we study, in which we consider all BH mergers in \Fable{} with $M_\text{BH} > 10^6$~\Msun{} at the time of numerical merger, residing in sufficiently massive subhaloes. These delays precede any microphysical time delays arising from  unresolved small scale hardening processes and are on average of the order of a few Gyrs, extending to $\gtrsim 12$~Gyrs, a timespan over which host galaxies undergo significant morphological changes. 
    \item We argue that the BH separations at the time of numerical BH mergers suffer from the same numerical artefacts as the merger times themselves. Using these as initial separations for calculating sub-grid hardening timescales may lead to systematic errors. A more reliable, albeit conservative, alternative is to use the subhalo (galaxy) separations from the snapshot immediately before the subhaloes merge. These separations are, on average, two orders of magnitude larger than the BH separations, with clear implications for the length of microphysical time delays.
    \item The introduction of macrophysical time delays suppresses high-mass merger events. Although accretion during the delay pushes surviving mergers toward higher chirp masses, it does not fully compensate for the suppression of the most massive events. Consequently, \textit{LISA} will detect a higher percentage of SMBH mergers, leading to a more complete view of the merging SMBH population.
    \item The reduction in BH merger rates, specifically in the highest-mass bin, will directly affect the SGWB detected by PTAs, with a suppression observed across the full detectable frequency range. Our findings also illustrate how assumptions about SMBH hardening mechanisms shape the predicted SGWB signal.
    \item The large macrophysical time delays imply an increase in the number of observable dual AGN at later cosmic times. Moreover, these dual AGN are predicted to be fainter than those in the original \Fable{} simulation, consistent with the observed redshift evolution of Eddington fractions, and residing at larger separations. 
\end{itemize}

It is important to note that our method assumes that our substructure finding algorithm can accurately identify these substructures which host BH particles. Substructure finding algorithms may struggle to identify subhaloes in the central region of the host halo and rather associate the corresponding particles to the central subhalo \citep[for a recent study on this see][]{moreno_2025}. If the central subhalo and any of these neighbouring subhaloes host a BH, our method may assume an earlier host galaxy merger time if the \textsc{Subfind} algorithm is unable to resolve the two structures. Thus, using more accurate subhalo-finding algorithms might increase our macrophysical delays in certain cases. Furthermore, all substructure finding algorithms will struggle with subhalo identification for low mass objects close to the resolution limit of simulations. Unfortunately, these are precisely the objects that host low mass or seed BHs which recent \textit{JWST} observations are starting to probe, indicating that very high resolution simulations are needed to understand this interesting BH mass regime.

In conclusion, our findings demonstrate that numerical choices in BH modelling within cosmological simulations significantly impact the predicted properties of merging SMBH populations. In this work, we quantify this issue and present a first attempt at mitigating for incorrect BH positions using a post-processing approach. However, the accuracy with which cosmological simulations capture the interplay between central BHs and their host galaxies cannot be fully reproduced through post-processing alone, underscoring a fundamental limitation of this method. Therefore, while these simulations remain a powerful tool for predicting current and future GW and EM observations, our results highlight the pressing need for more precise modelling of both BH and BH binary orbits, both above and below the resolution limit of simulations.

\section*{Acknowledgements}
The authors would like to thank Leah Bigwood for allowing us to use data from the \Fable{} $(100 \, h^{-1} \, \text{cMpc})^3$ cosmological box in this work, and for all her valuable assistance in helping us understand this data. We also thank Stephen Taylor and Sophie Koudmani for valuable comments on the paper. 

SB is supported by the Cambridge Centre for Doctoral Training in Data-Intensive Science funded by the UK Science and Technology Facilities Council (STFC). DS and MAB acknowledge support from the STFC, grant code ST/W000997/1. MAB additionally acknowledges support from a UKRI Stephen Hawking Fellowship (EP/X04257X/1). This work performed using the following facilities: the Cambridge Service for Data Driven Discovery (CSD3) operated by the University of Cambridge Research Computing Service (www.csd3.cam.ac.uk), provided by Dell EMC and Intel using Tier2 funding from the Engineering and Physical Sciences Research Council (capital grant EP/P020259/1), and DiRAC funding from the STFC (www.dirac.ac.uk), as well as the DiRAC@Durham facility operated by the Institute for Computational Cosmology on behalf of the STFC DiRAC HPC Facility (www.dirac.ac.uk). 
The equipment was funded by BEIS capital funding via STFC capital grants ST/P002293/1, ST/R002371/1 and ST/S002502/1, Durham University and STFC operations grant ST/R000832/1. DiRAC is part of the National e-Infrastructure.

\section*{Data Availability}

The \Fable{} data used in this paper is currently not publicly available through a web server, but is available upon request. All the scripts used to obtain the results in this work can be found at this \href{https://github.com/steph-buttigieg/BHMacroDelay}{GitHub repository} and with (very minor) modification can be run on the Illustris and IllustrisTNG data which are already publicly available.



\bibliographystyle{mnras}
\bibliography{references} 



\appendix

\section{Redshift evolution of the BH mass - stellar mass scaling relation} \label{app:evolution}

\begin{figure*}
    \centering
    \includegraphics[width = 2\columnwidth]{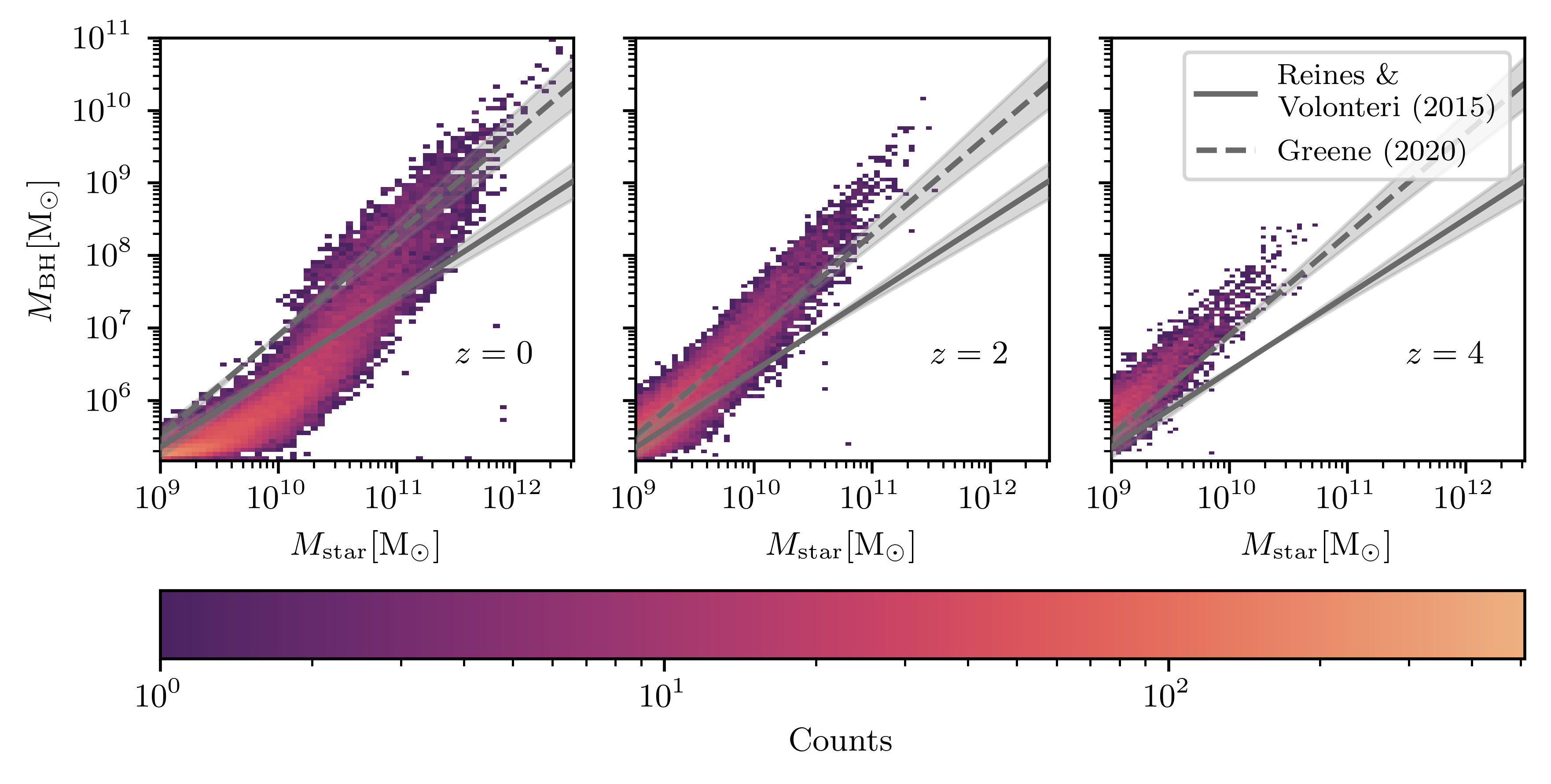}
    \caption{The evolution of the central BH mass vs. stellar mass within twice the half-mass radius of all galaxies in \Fable{}. The redshifts shown are $z = 0, 2, 4$. The best fits from \citet{reines_2015} and \citet{greene_2020} at $z = 0$ are also shown for comparison.}
    \label{fig:relation_evolution}
\end{figure*}

In Fig.~\ref{fig:relation_evolution}, we show the redshift evolution of the BH mass relation, where we plot the BH mass as a function of the stellar mass within twice the stellar half-mass radius of the host galaxy. We do this for easier comparison to observations, rather than using a proxy for the bulge mass as was done in Fig.~\ref{fig:mass-relation}, as it is increasingly more challenging to separate the mass of the bulge from the total stellar mass of galaxies at high redshifts~\citep{Davari_2016}. We compare these to observations by \citet{reines_2015} and \citet{greene_2020} at $z=0$. The cosmic evolution of this relation is not yet well understood, with recent \textit{JWST} measurements portraying a complex picture~\citep{Maiolino_2024}, and thus it is difficult to compare the population of BHs in \Fable{} to the observed population of SMBHs in this respect~\citep{Merloni_2009}. However, this figure indicates that host galaxies in the simulation are growing more rapidly than their central BHs, especially at the low mass end. Future observations with \textit{JWST} and GRAVITY+ will provide better constraints on this mass relation up to $z \gtrsim 2$, improving our understanding of black hole and galaxy co-evolution in the early universe~\citep{gravity_2022, Abuter_2024}.

\bsp	
\label{lastpage}
\end{document}